\documentclass[10pt,journal,final,letterpaper,oneside,twocolumn]{IEEEtran}

\usepackage{hyperref}
\usepackage{mathtools}
\usepackage{tcolorbox}
\usepackage{color}
\usepackage{theorem}
\usepackage{times,amsmath,epsfig}
\usepackage{amssymb}
\usepackage{subfigure}
\usepackage{cite}
\usepackage{cases}
\usepackage{siunitx}

\usepackage{algorithmic}
\usepackage{algorithm}
\usepackage{needspace}

\usepackage{tikz}
\usetikzlibrary{shapes,arrows}
\input{mysymbol.sty}
\tikzstyle{phantom vertex} = [ ellipse, 
                               anchor = center, 
                               minimum height = 1*\unit, 
                               minimum width  = 1*\unit,
                               inner sep=0pt,
                               anchor=center]
\tikzstyle{red vertex}   = [black, fill = red!20,   phantom vertex, draw]
\tikzstyle{black vertex} = [black, fill = black!20, phantom vertex, draw]
\tikzstyle{blue vertex}  = [black, fill = blue!20,  phantom vertex, draw]
\tikzstyle{green vertex} = [black, fill = green!20,  phantom vertex, draw]
\tikzstyle{yellow vertex} = [black, fill = yellow!20,  phantom vertex, draw]
\tikzstyle{cyan vertex} = [black, fill = cyan!20,  phantom vertex, draw]
\tikzstyle{vertex}       = [draw, phantom vertex]

\tikzstyle{point} = [ellipse, inner sep=0pt, draw, fill=white, anchor = center,
                     minimum height = 0.05*\unit, minimum width  = 0.05*\unit]

\newtheorem{mytheorem}{\bf Theorem}
\newtheorem{mydefinition}{\bf Definition}

\newtheorem{myproposition}{\bf Proposition}
\newtheorem{remark}{\bf Remark}

\newtcolorbox{myblockt}[1]{colback=urblue!5!white,
	colframe=urblue,fonttitle=\bfseries,
	title=#1}

\newtcolorbox{myblock}{colback=urblue!5!white,
	colframe=urblue,fonttitle=\bfseries}

\title{Unfolding WMMSE using Graph Neural Networks for Efficient Power Allocation}

\author{\IEEEauthorblockN{Arindam Chowdhury, Gunjan Verma, Chirag Rao, Ananthram Swami, and Santiago Segarra}
	\thanks{A. Chowdhury and S. Segarra are with the Dept. of ECE, Rice University.
		G. Verma, C. Rao, and A. Swami are with the US Army’s CCDC Army Research Laboratory.
		Research was sponsored by the Army Research Office and was accomplished under Cooperative Agreement Number W911NF-19-2-0269. 
		The views and conclusions contained in this document are those of the authors and should not be interpreted as representing the official policies, either expressed or implied, of the Army Research Office or the U.S. Government. 
		The U.S. Government is authorized to reproduce and distribute reprints for Government purposes notwithstanding any copyright notation herein.
		\newline
		Emails:  \{ac131, segarra\}@rice.edu, \hspace{1mm} \{gunjan.verma.civ, chirag.r.rao.civ, ananthram.swami.civ\}@mail.mil.}
}

\begin{document}
\maketitle

\begin{abstract}%
	We study the problem of optimal power allocation in a single-hop ad hoc wireless network.
	In solving this problem, we depart from classical purely model-based approaches and propose a hybrid method that retains key modeling elements in conjunction with data-driven components.
	More precisely, we put forth a neural network architecture inspired by the algorithmic unfolding of the iterative weighted minimum mean squared error (WMMSE) method, that we denote by unfolded WMMSE (UWMMSE).
	The learnable weights within UWMMSE are parameterized using graph neural networks (GNNs), where the time-varying underlying graphs are given by the fading interference coefficients in the wireless network. 
	These GNNs are trained through a gradient descent approach based on multiple instances of the power allocation problem.
	We show that the proposed architecture is permutation equivariant, thus facilitating generalizability across network topologies.
	Comprehensive numerical experiments illustrate the performance attained by UWMMSE along with its robustness to hyper-parameter selection and generalizability to unseen scenarios such as different network densities and network sizes.
 \end{abstract}


%
\section{Introduction}\label{S:Introduction}

Power and bandwidth are fundamental resources in communication, playing a key role in determining the effective capacity of a wireless channel~\cite{shannon1948mathematical,cave2007essentials}. 
In modern wireless communication systems, the scope of resources has been broadened to include beams in a multiple-input-multiple-output (MIMO) system, time slots in a time-division multiple access system (TDMA), frequency sub-bands in a frequency-division multiple access (FDMA) system, spreading codes in a code-division multiple access (CDMA) system, among several others~\cite{liang2019deep}. 
Optimal allocation of these resources under randomly varying channel characteristics and user demands is essential for the smooth operation of wireless systems. 
In particular, power allocation in a wireless ad hoc network is crucial to mitigate multi-user interference, one of the main performance-limiting factors.
In addition, transmission power of a mobile user is in itself a scarce resource. 
Indeed, careful and efficient power allocation is vital to prolong the battery life of mobile users~\cite{yu2016fair}.

A communication system should ensure fairness and quality of service to its users through efficient resource utilization.
Mathematically, the power allocation problem can be formulated as the problem of optimizing a certain system-level utility function -- such as sum-rate, min-rate, and harmonic-rate -- subject to resource budget constraints.
Despite the remarkable success of this paradigm~\cite{boche2011characterization}, it turns out that many of the formulated optimization problems are non-convex and NP-hard, making them difficult to solve~\cite{luo_2008_dynamic,liu2011coordinated,razaviyayn2013linear}.
Hence, a plethora of classical methods have been developed to obtain approximate solutions to these challenging optimization problems. 
For the canonical case of sum-rate maximization, the algorithms developed over the past decades can be broadly categorized into those based on Lagrangian dual decomposition~\cite{yu2006dual,hayashi2009spectrum}, interference pricing~\cite{huang2006distributed,wang2008price}, successive convex approximation~\cite{papandriopoulos2009scale}, and weighted minimum mean squared error (WMMSE) minimization \cite{shi2011iteratively}.
Of special interest to us due to its widespread adoption is this last category of classical methods.
Common shortcomings of this body of work include high computational complexity as well as the need of an accurate system model, which may not be feasible to construct in this era of constantly varying communication standards and user demands. High computational complexity has two particularly adverse consequences in wireless networks: 1) it more rapidly depletes device battery (which runs counter to the entire point of performing efficient power allocation), and 2) the power allocation needs to be recomputed frequently in time-varying channels, undermining the practical utility of a complex algorithm's time-delayed output that lags current channel conditions.
This has led the research community to look for faster, data-driven solutions for power allocation by leveraging machine learning tools.

Deep learning based methods have outperformed conventional approaches for wireless resource allocation in recent works \cite{lee2018deep, wang2018deep, nasir2019multi, zhang2019deep, chang2018learn, qin2019deep,chen2017machine}. 
Broadly, there are two ways of incorporating deep learning in solving resource allocation problems. 
One way is to follow a \emph{supervised} or \emph{imitation} learning approach where the functional approximation capacity of deep neural networks is leveraged to mimic established classical methods from solved instances~\cite{zhang2019energy}.
For example, a deep network has been employed~\cite{sun2018learning} to approximate the classical WMMSE solution~\cite{shi2011iteratively} for the power allocation problem in an interference channel. 
The main advantage of this paradigm is computational efficiency since, once trained, deep networks can run faster than the classical algorithms that they are imitating. 
By contrast, the main drawback is the poor generalization to scenarios (channel characteristics, number of users, user mobility) not represented in the original training examples. While expanding the training set to be more inclusive of the diversity of such phenomena is desirable, it may be infeasible to obtain groundtruth for such a large and representative set as that would involve repeated application of the computationally heavy iterative algorithm.
The second approach is \emph{unsupervised} where one parameterizes the power allocation function using a neural network and directly employs the optimization objective as a loss function, bypassing the need for solved problem instances~\cite{lee2018deep,eisen2019learning,eisen2020optimal,shen2019graph,meng2020power}. 
This loss function is minimized during training with respect to the parameters defining the neural network.
While such a procedure is computationally simple, no prior knowledge (e.g., based on classical optimization techniques) is leveraged to inform the algorithm's architectural or hyper-parameter choices.

In this paper, we advocate a third direction, which falls between the two described extremes and aims to address their shortcomings. 
The proposed method is \emph{unsupervised} in the sense that no solved instances of the power allocation problem are needed for training.
However, our approach still \emph{imitates} classical methods by incorporating part of their structure into the layered architecture of the neural network.
The goal is to leverage the theoretical models developed with expert knowledge and achieve near-optimal performance with significantly reduced execution time.
To accomplish this, we follow the paradigm of algorithm unfolding \cite{monga2019algorithm}.

Algorithm unfolding (or unrolling) was first proposed as an application of a parametric function to approximate the iterative soft-thresholding algorithm~\cite{gregor2010learning}. The idea is to unfold the iterations as a cascade of layers, where each layer has the same update structure as the original algorithm but the parameters are now learned from data. This approach was later successfully extended to several iterative algorithms that solve problems like blind deblurring~\cite{li2019algorithm}, non-negative matrix factorization~\cite{hershey2014deep}, semantic image segmentation \cite{liu2017deep}, clutter suppression~\cite{solomon2019deep}, rain removal~\cite{ding2018domain} and source separation~\cite{wang2018end}.
A standard practice in the above frameworks is to parameterize the function of interest using multi-layer perceptrons (MLPs) or convolutional neural networks (CNNs)~\cite{guo2019structure}. 
However, MLPs and CNNs -- which have proven to be tremendously effective in processing images, texts, and audio -- are not quite suitable for problems in wireless communication. 
In particular, the performance of these methods degrades dramatically when the network size becomes large since MLPs and CNNs fail to exploit the underlying topology of wireless networks.
This has led to several approaches that tried to adjust CNNs to the wireless setting \cite{lee2018deep,xu2019energy,van2019sum} capitalizing on, e.g., the spatial disposition of wireless networks \cite{cui2019spatial}.
We adopt an alternative direction~\cite{eisen2020optimal, nakashima2019deep, shen2019graph}, where graph neural networks (GNNs) \cite{bruna2013spectral,kipf2016semi,defferrard2016convolutional,li2017diffusion,gama2018convolutional,roddenberry_2019_hodgenet,yang_2018_enhancing} are used to parameterize the power allocation function, thus leveraging the natural representation of wireless networks as graphs.
GNNs utilize the structural relationships between nodes to locally process instantaneous channel state information.

In this context, we propose an unfolded weighted minimum mean squared error (UWMMSE) method, which is to the best of our knowledge the first GNN based deep unfolded architecture based on the iterative WMMSE algorithm. UWMMSE achieves state-of-the-art performance in utility maximization and computational efficiency for power allocation in wireless networks.

\medskip\noindent\textbf{Contribution.} 
The contributions of this paper are threefold:\\
i) We propose an unfolded version of WMMSE for power allocation in wireless networks -- where the learnable modules are parameterized via GNNs -- along with guidelines for its distributed implementation. Moreover, the modular approach advocated here can be followed as a roadmap for the derivation of other unfolded versions of WMMSE.\\
ii) In terms of theoretical analysis, we show that the permutation equivariance of the learning components implies permutation equivariance of the whole architecture; see Propositions~\ref{P:equiv} and~\ref{P:single_psi}. Additionally, we determine a necessary condition to be satisfied by the learning parameters to enable effective learning; see Theorem~\ref{T:convergence_necessary}. \\
iii) Through extensive numerical experimentation, we illustrate the performance of the proposed method compared with state-of-the-art alternatives, the impact on performance of the choice of key hyper-parameters in the determination of our architecture, and the generalizability to wireless networks of unseen sizes and densities.

\medskip\noindent\textbf{Paper outline.} 
In Section~\ref{S:Modeling}, we introduce the network topology and channel model that will be studied along with a formal statement of the power allocation problem.
After introducing the classical WMMSE algorithm, in Section~\ref{S:uwmmse} we present the proposed unfolded architecture; see Fig.~\ref{F:unr}.
Theoretical analyses of permutation equivariance and convergence are presented in Sections~\ref{Ss:gnns} and~\ref{Ss:convergence}, respectively.
We close this main section with a discussion on the distributed implementation and computational complexity of the proposed architecture in Section~\ref{Ss:distributed}.
Comprehensive numerical experiments in Section~\ref{S:num_exp} demonstrate the behavior of UWMMSE in comparison with established and recently proposed methods as well as for varying hyper-parameter and wireless network settings.
Conclusions and avenues for future research in Section~\ref{S:Conclusions} close the paper.

\medskip\noindent
\textbf{Notation.}
The entries of a matrix $\mathbf{X}$ and a vector $\mathbf{x}$ are denoted by $X_{ij}$ and $x_i$, respectively;
to avoid confusion, the alternative notation $[\mathbf{X}]_{ij}$ and $[\mathbf{x}]_{i}$ will be used occasionally.
Operations $(\cdot)^{\top}$ and $\mathbb{E}(\cdot)$ represent transpose and expected value, respectively.
$\mathbf{0}$ and $\mathbf{1}$ refer to the all-zeros and the all-ones vectors, respectively, where the sizes are clear from context.
$\diag(\bbX)$ denotes a diagonal matrix which stores the diagonal elements of $\bbX$. $\ccalN(0, \sigma^2)$ represents a zero-mean normal distribution with variance $\sigma^2$.

\section{System model and problem formulation}\label{S:Modeling}

We consider a single-hop ad hoc interference network having $M$ distinct single-antenna transmitter-receiver pairs. 
Transmitters are denoted by $i$ and the $i^{th}$ transmitter is associated to a single receiver denoted by $r(i)$ for $i= 1, \ldots, M$.  
Further, denoting the signal transmitted by $i$ as $x_i \in \mathbb{R}$, the received signal $y_i \in \mathbb{R}$ at $r(i)$ is given by
\begin{equation}\label{E:trans_model}
    y_i = h_{ii}x_i + \sum_{\substack{j=1 \, | \,  j\neq i}}^M h_{ij}x_j + n_i,
\end{equation}
where $h_{ii} \in \mathbb{R}$ is the channel between the $i^{th}$ transceiver pair, $h_{ij} \in \mathbb{R}$ for $i \neq j$ represents interference between transmitter $j$ and receiver $r(i)$, and $n_i \sim \ccalN(0, \sigma^2)$ for $\sigma > 0$ represents the additive channel noise. To accommodate the effect of fading channels, we slot time into discrete intervals and consider time-varying channels $h_{ij}(t)$, which we conveniently represent as a channel-state matrix $\bbH(t) \in \reals^{M \times M}$ where $[\bbH(t)]_{ij} = h_{ij}(t)$.
It will be instrumental to interpret $\bbH(t)$ as the time-varying and weighted adjacency matrix of a directed graph with $M$ nodes, where node $i$ represents the transmitter-receiver pair composed of transmitter $i$ and receiver $r(i)$. 

The instantaneous data rate $c_i$ achievable at receiver $r(i)$ is given by Shannon's capacity theorem as a function of the signal-to-interference-plus-noise-ratio (SINR), 
\begin{equation}\label{E:data_rate}
    c_i(\bbp(t), \bbH(t)) = \log_2 \left( 1 + \frac{| h_{ii} (t) |^2 p_i(t)}{\sigma^2 + \sum_{j\neq i} |h_{ij}(t)|^2 p_j(t)} \right),
\end{equation}
where $p_i(t) \in \mathbb{R}$, $p_i(t) \ge 0$, represents the power allocated to transmitter $i$ at time $t$ and $\bbp(t) = [p_1(t), \ldots, p_M(t)]^\top$.
Our objective is to determine the instantaneous power allocation vector $\bbp(t)$ that maximizes a network utility related to the data rates $c_i$.
Under the common understanding that channel states vary with time, from this point onward we omit the explicit dependence on $t$ to simplify notation. 

We formally state our power allocation problem as follows
\begin{align}\label{E:optimization_problem}
		 & \max_{\bbp} \,\, \sum_{i=1}^M \beta_i (c_i(\bbp, \bbH)) \quad  \\& \text{s.t.} \,\,\,\,\,  0 \leq p_i \leq p_{\max}, \,\,\,\, \text{for all} \,\, i, \nonumber
\end{align}
where $\beta_i$ is a generic increasing utility function of the data rate $c_i$ [cf.~\eqref{E:data_rate}], and $p_{\max}$ denotes the uniform maximum available power at every transmitter.
Notice that when selecting $\beta_i(z) = \alpha_i z$ for $\alpha_i>0$, the objective function in~\eqref{E:optimization_problem} boils down to the well-established (weighted) sum-rate utility. 
Moreover, other frequently encountered utilities are also encompassed by the formulation in~\eqref{E:optimization_problem} including proportional fairness $\beta_i(z) = \log(z)$ and harmonic mean rate $\beta_i(z) = -z^{-1}$.
However, even for well-established utility functions, the optimization problem in~\eqref{E:optimization_problem} has been shown to be NP-hard~\cite{luo_2008_dynamic, hong_2014_signal}. 
In this context, the present work is driven by the following question:

\vspace{1mm}

\begin{center}
How can we achieve an \emph{effective}, \emph{distributed}, and \emph{efficient} solution to~\eqref{E:optimization_problem}, where $\bbH$ is drawn from an \emph{accessible} distribution $\ccalH$?
\end{center}

\vspace{1mm}

In the above statement, by \emph{effective} we mean a solution that achieves comparable performance to a near-optimal classical approach, while \emph{accessible} distribution means that either $\ccalH$ is known or we can easily sample from it.

Several approximate methods have been proposed that aim to solve a tractable surrogate of~\eqref{E:optimization_problem} and, thence, find at least a good local maximum of the utility of interest. 
This classical body of work focuses on solving a single instance of~\eqref{E:optimization_problem} for an arbitrary $\bbH$, which must then be repeated to recompute the power allocation in successive time instants.

Given that in practice we are interested in solving several instances of~\eqref{E:optimization_problem} across time, a learning-based body of work has gained traction in the past years. 
In a nutshell, based on many channel state instances, the idea is to learn a map (i.e., a function approximator) between the channel state matrix $\bbH$ and the corresponding (approximate) optimal power allocation $\bbp$. 
In this way, when a new channel is drawn, the power can be efficiently computed using the learned map without the need for solving an optimization problem. Unlike common neural-network based functional approximators whose inputs and outputs are often of fixed dimensionality, a distinguishing property of the problem here is that $\bbH$'s dimension is not necessarily fixed a-priori and can indeed change as nodes enter and exit the wireless network. This fact combined with the inherent topological structure present in $\bbH$ foreshadows the importance of using graph neural networks as the functional approximator, a point we further elaborate in Section~\ref{S:uwmmse}.

Our goal is to combine the advantages of the classical  and learning-based paradigms.
More specifically, we seek to address our driving question by leveraging the \emph{approximate}, \emph{distributed} and \emph{interpretable} solution provided by the classical WMMSE method while enhancing it with the computational \emph{efficiency} of trained machine learning models. 
We pursue this synergistic combination under the paradigm of algorithm unfolding, as discussed in the next section.

\section{Unfolded weighted minimum mean squared error (UWMMSE)}\label{S:uwmmse}

\begin{figure*}
	\centering
	\includegraphics[width=0.75\linewidth]{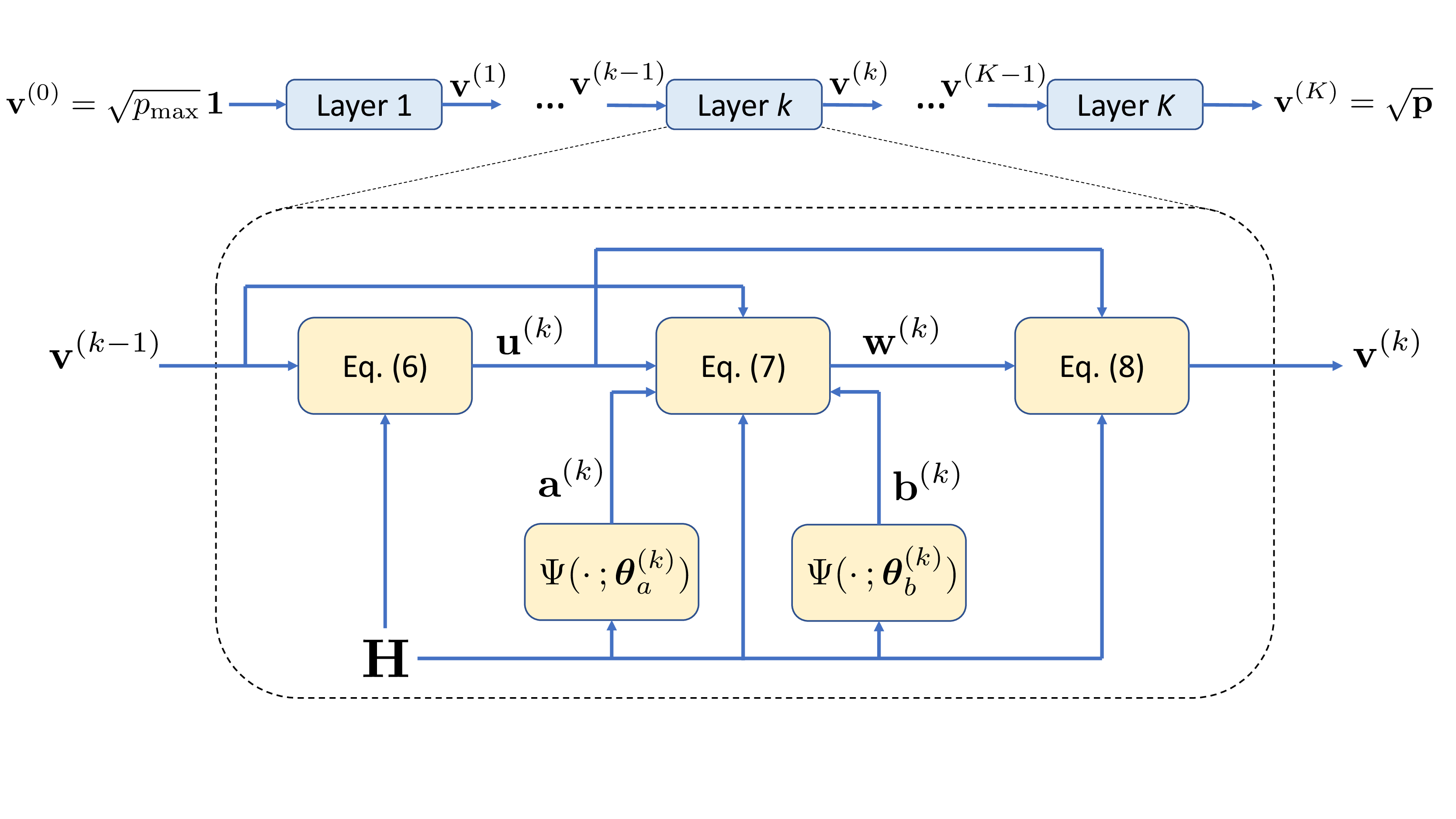}
	\vspace{-4mm}
	\caption{\small Schematic diagram of the proposed unfolded weighted minimum mean squared error (UWMMSE) algorithm.
	The input to the layered structure is $\bbv^{(0)} = \sqrt{p_{\max}} \, \mathbf{1}$ and the computed power allocation is given by the elementwise square of the output $\bbv^{(K)}$. 
	A generic intermediate layer $k$ is detailed. 
	The five equations in \eqref{E:unfold_1}-\eqref{E:unfold_4} are represented by the yellow blocks and the dependence on intermediate variables is made explicit in the flow diagram. 
	The parameters $\bbtheta_a^{(k)}$ and $\bbtheta_b^{(k)}$ for all layers $k$ are learned to minimize the loss $\ell(\bbTheta)$ in \eqref{E:loss_sgd}, thus promoting a faster convergence than its classical WMMSE counterpart.}
	\vspace{-4mm}
	\label{F:unr}
\end{figure*}

Algorithm unfolding (or unrolling)~\cite{balatsoukas2019deep,liu2019deep,monga2019algorithm,farsad2020data} refers to the general notion of selecting an iterative model-based approach to solve a problem of interest and building a problem-specific neural network with layers inspired by these iterations.
Backed-up by its recent success on several problems, we present here the first unfolded algorithm for power allocation in wireless networks.
Notice, however, that the action of unfolding is not uniquely determined. 
Several decisions must be made with the objective of incorporating domain knowledge as well as striking the right balance between interpretability and performance.
These decisions include selecting what parameters become trainable and what trainable architectures are used to represent these parameters.
In this paper, we develop and evaluate a way to unfold WMMSE using graph neural networks. 

Before presenting our unfolded architecture, we introduce the basics of WMMSE~\cite{shi2011iteratively}.
Essential for this classical algorithm is to reformulate~\eqref{E:optimization_problem} as
\begin{align}\label{E:problem_reformulation}
&\min_{\mathbf{w,u,v}} \sum_{i=1}^M (w_i e_i - \log w_i ),\\
& \text{st.} \quad e_i = (1-u_i h_{ii} v_i)^2 + \sigma^2 u_i^2 + \sum_{i \neq j} u_i^2 h_{ij}^2 v_j^2, \nonumber \\
& \qquad \,\, \, v^2_i \leq p_{\max}, \,\,\,\, \text{for all} \,\, i, \nonumber
\end{align}
where $e_i$ computes the mean-square error of the signal at node $i$ under the assumption that the transmitted signal $x_i$ is independent from the noise $n_i$ [cf.~\eqref{E:trans_model}]. For simplicity, in~\eqref{E:problem_reformulation} we have focused on the sum-rate utility function ($\beta_i(z) = z$ for all $i$). 
However, the developments here presented can be extended to more general utilities $\beta_i$; see Remark~\ref{R:different_utilities}.

We say that~\eqref{E:problem_reformulation} is equivalent to~\eqref{E:optimization_problem} because it can be shown~\cite[Thm. 3]{shi2011iteratively} that the optimal solution $\{\bbw^*, \bbu^*, \bbv^*\}$ of the former and that of the latter $\bbp^*$ are related as $\sqrt{\bbp^*} = \bbv^*$, where the square root is applied elementwise.  
Hence, if we obtain a optimal solution to the (non-convex) problem in~\eqref{E:problem_reformulation}, this would immediately translate into a optimal solution to our problem of interest~\eqref{E:optimization_problem}. 
Additionally, the structure in~\eqref{E:problem_reformulation} is amenable to a block-coordinate descent solution, since the objective is convex in each of the three variables when the other two are fixed.
Indeed, there exist closed-form expressions for these coordinate descent steps.
Our unfolding architecture is inspired by these iteratively applied closed-form solutions, which we augment with learnable parameters.

We propose to compute the allocated power as a function of the channel state matrix $\bbp = \Phi(\bbH; \bbTheta)$ through a layered architecture $\Phi$ with trainable weights $\bbTheta$.

More precisely, setting $\bbv^{(0)} = \sqrt{p_{\max}} \, \mathbf{1}$, we have that for layers $k = 1, \ldots, K$, 
\begin{align}
\bba^{(k)} &= \Psi(\bbH; \bbtheta_a^{(k)}),  \qquad \bbb^{(k)}  = \Psi(\bbH;\bbtheta_b^{(k)}), \hspace{-10mm} \label{E:unfold_1}\\
u^{(k)}_i &= \frac{h_{ii}v^{(k-1)}_i}{\sigma^2 + \sum_j h_{ij}^2 {v^{(k-1)}_j} v^{(k-1)}_j}, \,\, &&\text{for all } i, \label{E:unfold_2}\\
w^{(k)}_i &= \frac{a_i^{(k)}}{1 - u^{(k)}_i h_{ii} v^{(k-1)}_i} + b_i^{(k)}, &&\text{for all } i, \label{E:unfold_3}\\
v^{(k)}_i &= \alpha \left( \frac{ u^{(k)}_i h_{ii} w^{(k)}_i}{\sum_j  h_{ji}^2 u^{(k)}_j u^{(k)}_j w^{(k)}_j}\right), &&\text{for all } i,\label{E:unfold_4}
\end{align}
and the output power is determined as $\bbp = \Phi(\bbH; \bbTheta) = (\bbv^{(K)})^2$, where the square is applied elementwise.
The non-linear function $\alpha(z) \coloneqq [z]_0^{\sqrt{p_{\max}}}$ in~\eqref{E:unfold_4} simply ensures that $v_i^{(k)} \in [0, \sqrt{p_{\max}}]$ by saturating the right-hand side of~\eqref{E:unfold_4} at these extreme values. 
This guarantees that the constraint in~\eqref{E:optimization_problem} is satisfied.
The trainable parameters $\bbTheta$ are given by the concatenation of $\bbtheta_a^{(k)}$ and $\bbtheta_b^{(k)}$ in~\eqref{E:unfold_1} for all layers, i.e., $\bbTheta = [\bbtheta_a^{(1)}, \bbtheta_b^{(1)}, \ldots, \bbtheta_a^{(K)}, \bbtheta_b^{(K)}]$. 
Finally, the functions $\Psi$ parametrized by $\bbTheta$ in~\eqref{E:unfold_1} are chosen to be graph neural networks, as further discussed in Section~\ref{Ss:gnns}. 
A schematic view of our proposed layered architecture is presented in Fig.~\ref{F:unr}.

To better understand the mechanics of the proposed layered architecture, notice that each layer $k$ as described in \eqref{E:unfold_1}-\eqref{E:unfold_4} is characterized by five vectors $\bba^{(k)}, \bbb^{(k)}, \bbu^{(k)}, \bbw^{(k)}, \bbv^{(k)} \in \reals^M$. 
If we set $\bba^{(k)} = \mathbf{1}$ and $\bbb^{(k)} = \mathbf{0}$, then expressions \eqref{E:unfold_2}-\eqref{E:unfold_4} correspond to the closed-form solutions of a block-coordinate algorithm designed to solve~\eqref{E:problem_reformulation}~\cite{shi2011iteratively}. Putting it differently, for these values of $\bba^{(k)}$ and $\bbb^{(k)}$, our proposed architecture boils down to the classical WMMSE truncated to $K$ iterations. 
In this setting, $\bbu^{(k)}$ and $\bbv^{(k)}$ represent receiver and transmitter side variables, respectively, in the sense that $u_i^{(k)}$ depends exclusively on the channel states \emph{into} receiver $r(i)$ whereas $v_i^{(k)}$ depends exclusively on the channel states \emph{out of} transmitter $i$. 
Moreover, upon convergence, these represent measures of the strength of signals at the receiver and transmitter for our single-input single-output (SISO) system. 
Their equivalent variables in the MIMO case would correspond to receiver and transmitter-side beamformers; see~\cite{shi2011iteratively} for additional details.
Notice also that the update of $\bbw^{(k)}$ in \eqref{E:unfold_3} is local in the sense that $w_i^{(k)}$ only depends on the $i$th entries of the other vectors of interest and the channel state $h_{ii}$ between transmitter $i$ and its intended receiver $r(i)$. 

In spite of being the default algorithm for optimal power allocation, WMMSE entails high computational and time complexity. This complexity arises because WMMSE requires many iterations of the updates \eqref{E:unfold_2}-\eqref{E:unfold_4} for convergence. 
Hence, the objective of the newly introduced variables $\bba^{(k)}$ and $\bbb^{(k)}$ is to accelerate this convergence while maintaining a good performance.
Intuitively, if we learn a smarter update rule for $\bbw^{(k)}$ that accelerates its convergence, we can achieve good performance with only a few iterations of WMMSE. 
In this sense, our proposed architecture is an unfolded version of the classical algorithm.
Notice that additional learning parameters and more sophisticated functional forms could be included in the updates~\eqref{E:unfold_2}-\eqref{E:unfold_4}. However, the simple learned affine transformation proposed in~\eqref{E:unfold_3} achieves good performance in practice (see Section~\ref{S:num_exp}) while being simple to implement.

Under the natural assumption that the weights $\bba^{(k)}$ and $\bbb^{(k)}$ should depend on the channel state $\bbH$, we advocate a learning-based method where this dependence is made explicit via the parametric functions $\Psi$ in~\eqref{E:unfold_1}. 
For fixed parameters $\bbTheta$, the allocated power for a channel state $\bbH$ is given by $\Phi(\bbH; \bbTheta)$ and results in a sum-rate utility of $\sum_{i=1}^M c_i(\Phi(\bbH; \bbTheta), \bbH)$; [cf.~\eqref{E:optimization_problem} for $\beta_i(z) = z$]. 
Hence, we define the loss function
\begin{equation}\label{E:loss_sgd}
\ell(\bbTheta) = - \mathbb{E}_{\bbH \sim \ccalH} \left[ \sum_{i=1}^M c_i(\Phi(\bbH; \bbTheta), \bbH) \right].
\end{equation}
Even if $\ccalH$ is known, minimizing $\ell(\bbTheta)$ with respect to $\bbTheta$ entails solving a non-convex optimization problem. 
Under the assumption that the function $\Psi(\cdot\, ; \bbtheta)$ in~\eqref{E:unfold_1} is differentiable with respect to $\bbtheta$ and leveraging the fact that we have access to samples of $\ccalH$ (cf. Section~\ref{S:Modeling}), we seek to minimize~\eqref{E:loss_sgd} through stochastic gradient descent.
It should be noted that, unlike~\cite{sun2018learning} and in accordance with some recent works~\cite{shen2019graph, eisen2020optimal}, UWMMSE is an \emph{unsupervised method} since it requires access to samples of the channel state matrices $\bbH$ but does \emph{not} require access to the optimal power allocations (labels) associated with those channels.

\begin{remark}[Different utility functions]\label{R:different_utilities}
	\normalfont Our proposed UWMMSE is well suited to solve the optimal power allocation problem for objective functions that go beyond the sum-rate utility.
	Adopting the notation of a generic utility $\beta_i$ in \eqref{E:optimization_problem}, UWMMSE can be adapted to this more encompassing setting by incorporating two changes.
	First, the loss function in \eqref{E:loss_sgd} must be updated to $\ell(\bbTheta) = - \mathbb{E}_{\bbH \sim \ccalH} \left[ \sum_{i=1}^M \beta_i(c_i(\Phi(\bbH; \bbTheta), \bbH)) \right]$.
	Second, and more noteworthy, we must adapt the unfolded architecture to correspond to the WMMSE solution of the modified utility problem. 
	This is unlike state-of-the-art methods~\cite{sun2018learning, shen2019graph, eisen2020optimal} where the architecture is unaffected by the utility at hand, thus indicating a limited incorporation of domain knowledge. 
	More precisely, defining the scalar functions $\gamma_i(z) = - \beta_i (-\log(z))$, the update in \eqref{E:unfold_3} must be changed to
	\begin{equation}\label{E:generic_w}
	w_i^{(k)} = \gamma_i'(1 - u^{(k)}_i h_{ii} v^{(k-1)}_i) \, a_i^{(k)} + b_i^{(k)},
	\end{equation} 
	where $\gamma' = \frac{d\gamma}{dz}$, while the remaining equations in \eqref{E:unfold_1}-\eqref{E:unfold_4} are left unaltered. 
	Two observations are in order. First, for the specific case of sum-rate utility, we have that $\beta_i(z) = z$. This implies that $\gamma_i(z) = \log(z)$ and it becomes apparent that \eqref{E:unfold_3} is recovered from~\eqref{E:generic_w}.
	Second, one can leverage~\cite[Thm. 2]{shi2011iteratively} to show that the optimal power allocation is a fixed point of our modified iteration for $a_i^{(k)} = 1, b_i^{(k)} =0$ under the condition that $\gamma_i(z)$ is strictly concave for $z>0$. 
	The empirical effect of the modification in~\eqref{E:generic_w} is studied in Section~\ref{Ss:node_feat_utility}.
\end{remark}


\subsection{Graph neural networks: Permutation equivariance}\label{Ss:gnns}

Thus far, we have been purposely nonspecific in the definition of the parametric function $\Psi(\cdot\,;\bbtheta)$ in~\eqref{E:unfold_1} in order to highlight the generality of the proposed unfolded algorithm. 
We have only identified differentiability with respect to $\bbtheta$ as a requirement to minimize~\eqref{E:loss_sgd} via stochastic gradient descent. 
In this section, we further winnow the set of candidate functions $\Psi$ by requiring \textit{permutation equivariance}. 
As will become apparent, \emph{graph neural networks} (GNNs) arise as natural candidates for the function $\Psi$.
GNNs are the natural extensions of convolutional neural networks to graph domains, where the architectures are appropriate to exploit non-Euclidean data.
In a nutshell, the classical convolutional layers are replaced by (banks of) graph filters~\cite{segarra_2017_optimal} that describe learnable local aggregation operations.

We introduce a formal definition for permutation equivariance. Consider the set $\ccalF$ of all functions $f: \reals^{M \times M} \to \reals^M$ and a generic permutation matrix $\bbPi \in \{0,1\}^{M \times M}$.

\begin{mydefinition}
	A function $f \in \ccalF$ is permutation equivariant if $f(\bbPi \bbH \bbPi^\top) = \bbPi f(\bbH)$ for all matrices $\bbH$ and all permutations $\bbPi$.
\end{mydefinition}

Intuitively, if we permute the labels of the nodes in our network before computing the permutation equivariant function $f$, the individual output values are not changed but only permuted by this same permutation. 
Notice that this is specially relevant in our context since the node indexing is arbitrary and should not play any role in the determination of the optimal power allocation.




Naturally, the choice of $\Psi$ in~\eqref{E:unfold_1} affects the permutation equivariance of our proposed method, as we state next.

\begin{myproposition}\label{P:equiv}
If $\Psi( \cdot\,; \bbtheta)$ in~\eqref{E:unfold_1} is permutation equivariant then the UWMMSE method $\Phi(\cdot\,; \bbTheta)$ is also permutation equivariant.
\end{myproposition}	
\begin{IEEEproof}
	Denote by $\bbv^{(k)} = \Phi_k(\bbH, \bbv^{(k-1)}; \bbtheta^{(k)})$ the output of the $k$th layer of the UWMMSE architecture in \eqref{E:unfold_1}-\eqref{E:unfold_4}. 
	Also, denote by $\tilde{\bbH} = \bbPi \bbH \bbPi^\top$ and $\tilde{\bbv}^{(k-1)} = \bbPi {\bbv}^{(k-1)}$ arbitrary permuted versions of the channel matrix and the input to the $k$th layer.
	We are first going to establish that $\Phi_k(\tilde{\bbH}, \tilde{\bbv}^{(k-1)}; \bbtheta^{(k)}) = \bbPi \bbv^{(k)}$. 
	From the permutation equivariance of $\Psi$ we have that $\tilde{\bba}^{(k)} = \Psi(\tilde{\bbH}; \bbtheta_a^{(k)}) = \bbPi {\bba}^{(k)}$ and, similarly, $\tilde{\bbb}^{(k)} = \bbPi {\bbb}^{(k)}$. 
	Denoting by $\pi(i)$ the new index of node $i$ upon permutation $\bbPi$, it follows from \eqref{E:unfold_2} that
	\begin{align*}
	\tilde{u}^{(k)}_i &= \frac{\tilde{h}_{ii}\tilde{v}^{(k-1)}_i}{\sigma^2 + \sum_j \tilde{h}_{ij}^2 {\tilde{v}^{(k-1)}_j} \tilde{v}^{(k-1)}_j} \\
	& = \frac{{h}_{\pi(i) \pi(i)} {v}^{(k-1)}_{\pi(i)}}{\sigma^2 + \sum_j {h}_{\pi(i) \pi(j)}^2 {{v}^{(k-1)}_{\pi(j)}} {v}^{(k-1)}_{\pi(j)}} = {u}^{(k)}_{\pi(i)},
	\end{align*}
	which, in vector form, equates to $\tilde{\bbu}^{(k)} = \bbPi {\bbu}^{(k)}$. 
	Similarly, from \eqref{E:unfold_3} and \eqref{E:unfold_4} it follows that $\tilde{\bbw}^{(k)} = \bbPi {\bbw}^{(k)}$ and $\tilde{\bbv}^{(k)} = \Phi_k(\tilde{\bbH}, \tilde{\bbv}^{(k-1)}; \bbtheta^{(k)})= \bbPi {\bbv}^{(k)}$, as wanted. 
	
	We are now going to leverage this to show that $\Phi(\cdot\,; \bbTheta)$ is equivariant. 
	Consider the case where $K=1$. From the definition of $\Phi(\cdot\,; \bbTheta)$ we obtain that
	\begin{align*}
	\Phi( \tilde{\bbH}; \bbTheta) &= \Phi_1(\tilde{\bbH}, \bbv^{(0)}; \bbtheta^{(1)}) = \Phi_1(\tilde{\bbH}, \bbPi \bbv^{(0)}; \bbtheta^{(1)}) \\
	&= \bbPi \Phi_1({\bbH}, \bbv^{(0)}; \bbtheta^{(1)}) = \bbPi \Phi({\bbH}; \bbTheta),
	\end{align*}
	where the second equality follows from the fact that $\bbv^{(0)}$ is a constant vector and the third equality is a special case of the identity previously established for $k=1$.
	This shows that a single layer UWMMSE is permutation equivariant. That permutation equivariance holds for $K>1$ can be shown via a simple induction argument here omitted.
\end{IEEEproof}

Having established that permutation equivariance is a desirable property for UWMMSE and guided by Proposition~\ref{P:equiv}, we proceed to specify a particular $\Psi$. Inspired by the simplicity of graph convolutional networks in~\cite{kipf2016semi}, we propose the following function
\begin{align}
\Psi(\bbH; \bbtheta) & = \alpha_2 \left( \mathrm{diag}(\bbH) \, \bbZ \, \bbomega_{21} + \bbH \, \bbZ \,\bbomega_{22} \right), \label{E:psi}\\
\bbZ & = \alpha_1 \left( \mathrm{diag}(\bbH) \, \mathbf{1} \, \bbomega^\top_{11} + \bbH \, \mathbf{1} \,\bbomega^\top_{12} \right), \label{E:psi_2}
\end{align}
where $\bbomega_{11}, \bbomega_{12}, \bbomega_{21}, \bbomega_{22} \in \reals^F$ for some predefined number of hidden features $F$, $\bbtheta = [\bbomega_{11}^\top, \bbomega_{12}^\top, \bbomega_{21}^\top, \bbomega_{22}^\top]^\top$, $\alpha_1$ is the rectified linear unit (ReLU) activation function and $\alpha_2$ is the sigmoid activation function.
The proposed function satisfies our desired property, as stated next.
\begin{myproposition}\label{P:single_psi}
	Function $\Psi( \cdot\,; \bbtheta)$ in~\eqref{E:psi} is permutation equivariant.
\end{myproposition}
\begin{IEEEproof}
Notice that
\begin{align*}
\alpha_1 \left( \mathrm{diag}(\bbPi \bbH \bbPi^\top) \, \mathbf{1} \, \bbomega^\top_{11} + \bbPi \bbH \bbPi^\top \, \mathbf{1} \,\bbomega^\top_{12} \right) & = \\
 \alpha_1 \left( \bbPi\mathrm{diag}( \bbH ) \, \mathbf{1} \, \bbomega^\top_{11} + \bbPi \bbH \, \mathbf{1} \,\bbomega^\top_{12} \right) & = \bbPi \bbZ,
\end{align*}
where we have used the facts that $\mathrm{diag}(\bbPi \bbH \bbPi^\top) = \bbPi \mathrm{diag}( \bbH ) \bbPi^\top$ and $\bbPi^\top \mathbf{1} = \mathbf{1}$. 
Replacing this in~\eqref{E:psi} we have that
\begin{align*}
&\Psi(\bbPi \bbH \bbPi^\top; \bbtheta) \\ 
& \,\,\,\, = \alpha_2 \left( \mathrm{diag}(\bbPi \bbH \bbPi^\top) \, \bbPi \bbZ \, \bbomega_{21} + \bbPi \bbH \bbPi^\top \, \bbPi \bbZ \,\bbomega_{22} \right)\\
& \,\,\,\, = \alpha_2 \left( \bbPi \mathrm{diag}(\bbH) \, \bbZ \, \bbomega_{21} + \bbPi \bbH \bbZ \,\bbomega_{22} \right) = \bbPi \Psi( \bbH; \bbtheta),
\end{align*}
as wanted. 
\end{IEEEproof}

From the combination of Propositions~\ref{P:equiv} and~\ref{P:single_psi} it immediately follows that the proposed UWMMSE is permutation equivariant.
Importantly, the modular description of UWMMSE put forth in~\eqref{E:unfold_1}-\eqref{E:unfold_4} allows the practitioner to choose their preferred GNN in defining $\Psi$. Moreover, Proposition~\ref{P:equiv} guarantees that, as long as the chosen $\Psi$ is permutation equivariant, this desirable feature will permeate to UWMMSE.
Henceforth, we adopt $\Psi$ as in~\eqref{E:psi} with the exception of Section~\ref{Ss:ablation} where the performance of different choices of $\Psi$ is compared. 

\begin{remark}[Incorporating node features]\label{R:node_features}
	\normalfont GNNs are well suited to leverage the structure of a given graph to process values defined on its nodes, often referred to as node features or graph signals \cite{kipf2016semi, marques_2016_sampling}.
	However, in our proposed implementation [cf.~\eqref{E:unfold_1} and~\eqref{E:psi}], there are no node features involved. 
	In the presence of node features, our proposed framework can be easily enriched to encompass this information. 
	More precisely, collecting $F'$ node features in the matrix $\bbQ \in \reals^{N \times F'}$ where $q_{if}$ stores the values of feature $f$ at node $i$, we may consider functions of the type $\Psi(\bbH, \bbQ; \bbtheta)$ in~\eqref{E:unfold_1}. 
	In particular, we still propose the GNN in~\eqref{E:psi} replacing $\mathbf{1}$ in~\eqref{E:psi_2} by $\bbQ$ and resizing the parameter matrices accordingly. The theory discussed in this section still holds, where the modified notion of permutation equivariance boils down to $\Phi(\bbPi \bbH \bbPi^\top, \bbPi \bbQ; \bbTheta) = \bbPi \Phi( \bbH, \bbQ; \bbTheta)$. 
	Examples of node features in the context of power allocation include relative node weights in the determination of a weighted sum-rate utility, queue lengths, traffic rates, and topological features of the underlying graph; see Section~\ref{Ss:node_feat_utility}  for an implementation of this enriched setup.
\end{remark}


\subsection{UWMMSE convergence: A necessary condition}\label{Ss:convergence}

To further study the proposed method, we analyze its behavior as the number of layers $K$ grows. 
Intuitively, since the optimal power allocation is a fixed point of the classical WMMSE iteration~\cite{shi2011iteratively}, one would expect deeper layers of the UWMMSE to more closely resemble its classical counterpart.
To formalize this intuition, let us introduce the following notation.
For any given channel state $\bbH$, let us define the optimal power allocation, i.e., the solution to~\eqref{E:optimization_problem}, as $\bbp^*$.
Accordingly, we define $v_i^* = \sqrt{p_i^*}$, $u_i^*$ as given by~\eqref{E:unfold_2} when $\bbv^{(k-1)}$ is replaced by $\bbv^*$, and $w_i^* = (u_i^* h_{ii} v_i^*)^{-1}$. 
Also, for simplicity we consider the case where $\bba^{(k)} = \mathbf{1}$ for all $k$ and study the convergence of $\bbb^{(k)}$.
With this notation in place, the following result can be shown.

\begin{mytheorem}\label{T:convergence_necessary}
	Consider a UWMMSE architecture~\eqref{E:unfold_1}-\eqref{E:unfold_4} of infinite depth with $\bba^{(k)} = \mathbf{1}$ for all $k$. If $\bbv^{(k)} \to \sqrt{\bbp^*}$ uniformly as $k \to \infty$ then, for all $i$ such that $0< p_i^* < p_{\max}$, it must hold that
	\begin{equation}\label{E:convergence_necessary_condition}
	\sum_{j \neq i} h_{ji}^2 (u_j^*)^2 (w_i^* b_j^{(k)} - w_j^* b_i^{(k)}) \to 0 \,\, \text{as} \,\, k \to \infty.
	\end{equation}
\end{mytheorem}
\begin{IEEEproof}
	Since $\bbv^{(k)} \to \sqrt{\bbp^*} = \bbv^*$ uniformly, for all $\eta > 0$ there must exist a layer index $K_1$ such that for all $k > K_1$, $v_i^{(k-1)} = v_i^* + \epsilon_i$ and $v_i^{(k)} = v_i^* + \epsilon'_i$ where $|\epsilon_i| < \eta$ and $|\epsilon'_i| < \eta$ for all $i$.
	Resorting to the notation used in the proof of Proposition~\ref{P:equiv}, we have that $\bbv^{(k)} = \Phi_k(\bbH, \bbv^{(k-1)}; \bbb^{(k)})$, where we have made explicit the dependence on $\bbb^{(k)}$. From the uniform convergence, it follows that
	\begin{equation}\label{E:proof_necessary_condition}
	\bbv^* + \bbepsilon' = \bbv^{(k)} = \Phi_k(\bbH, \bbv^* + \bbepsilon; \bbb^{(k)}).
	\end{equation}
	Notice that if we find an expression $g_k(\bbH, \bbv^* + \bbepsilon; \bbb^{(k)})$ such that $\Phi_k(\bbH, \bbv^* + \bbepsilon; \bbb^{(k)}) = \bbv^* + g_k(\bbH, \bbv^* + \bbepsilon; \bbb^{(k)})$, we might replace this in~\eqref{E:proof_necessary_condition} to obtain
	\begin{equation}\label{E:proof_necessary_condition_2}
	\bbepsilon' = g_k(\bbH, \bbv^* + \bbepsilon; \bbb^{(k)}).
	\end{equation}
	Since~\eqref{E:proof_necessary_condition_2} must hold for arbitrary small but positive $\eta$, this implies that $g_k(\bbH, \bbv^* + \bbepsilon; \bbb^{(k)}) \to 0$ as $k \to \infty$. In the remainder of the proof, we find an explicit form for $g_k(\bbH, \bbv^* + \bbepsilon; \bbb^{(k)})$ and show that this implies~\eqref{E:convergence_necessary_condition}. 
	
	We begin by replacing $\bbv^{(k-1)}$ by $\bbv^* + \bbepsilon$ in~\eqref{E:unfold_2} to obtain 
	\begin{align}\label{E:proof_necessary_condition_3}
	u^{(k)}_i & = \frac{h_{ii} (v_i^* + \epsilon_i)}{\sigma^2 + \sum_j h_{ij}^2 (v_j^* + \epsilon_j)^2}, \\
	& = \frac{h_{ii} v_i^* + h_{ii} \epsilon_i}{\sigma^2 + \sum_j h_{ij}^2 [(v_j^*)^2 + 2 v_j^* \epsilon_j + \epsilon^2_j]} \nonumber \\
	& = \frac{h_{ii} v_i^*}{\sigma^2 \!+\! \sum_j h_{ij}^2 (v_j^*)^2}  \!+\! \frac{h_{ii} \epsilon_i}{\sigma^2 + \sum_j h_{ij}^2 [(v_j^*)^2 + 2 v_j^* \epsilon_j + \epsilon^2_j]} \nonumber \\
	& -  \frac{h_{ii} v_i^*}{\sigma^2 + \sum_j h_{ij}^2 (v_j^*)^2} 
	\,\,  \frac{\sum_j h_{ij}^2 [2 v_j^* \epsilon_j + \epsilon^2_j]}{\sigma^2 + \sum_j h_{ij}^2 [(v_j^*)^2 + 2 v_j^* \epsilon_j + \epsilon^2_j]}  \nonumber \\
	& = u_i^* + \epsilon_{u_i}, \nonumber 
	\end{align}
	where the last equality follows from the definition of $u_i^*$ and $\epsilon_{u_i}$ contains all the terms depending on $\bbepsilon$ so that $|\epsilon_{u_i}| \to 0$ as $\eta \to 0$. We then move to~\eqref{E:unfold_3} and replace $u^{(k)}_i$, $v^{(k-1)}_i$, and $a^{(k)}_i$ by $u_i^* + \epsilon_{u_i}$, $v_i^* + \epsilon_{i}$ and $1$, respectively. Following a procedure similar to that in~\eqref{E:proof_necessary_condition_3} we get that 
	\begin{align}\label{E:proof_necessary_condition_4}
	w^{(k)}_i = w_i^* + \epsilon_{w_i} + b^{(k)}_i,
	\end{align}
	where $|\epsilon_{w_i}| \to 0$ as $\eta \to 0$.
	We repeat the procedure for~\eqref{E:unfold_4}, where we notice that, for large enough $k$, the non-linearity $\alpha$ is moot since $v_i^{(k)} \to \sqrt{p_i^*}$ and $0< p_i^* < p_{\max}$. 
	With this in mind, it follows from~\eqref{E:unfold_4} that
	\begin{align}\label{E:proof_necessary_condition_5}
	v^{(k)}_i &= \frac{ u^*_i h_{ii} w^*_i}{\sum_j  h_{ji}^2 (u^*_j)^2 w^*_j} + \epsilon_{v_i} \\ & + \frac{h_{ii} u^*_i}{\sum_j \! h_{ji}^2 (u^*_j)^2 w^*_j} \,
	 \frac{\sum_{j \neq i} h_{ji}^2 (u_j^*)^2 (w_j^* b_i^{(k)} - w_i^* b_j^{(k)})}{\sum_j h_{ji}^2 [(u_j^* + \epsilon_{u_j})^2(w^*_j + \epsilon_{w_j} \! + b_j^{(k)})] } 
	\end{align}
	where $|\epsilon_{v_i}| \to 0$ as $\eta \to 0$.
	From the fact that $(\bbu^*, \bbw^*, \bbv^*)$ is a fixed point of the classical WMMSE iteration \cite{shi2011iteratively}, it follows that the first term in the right-hand side of~\eqref{E:proof_necessary_condition_5} is equal to $v_i^*$. 
	This implies that the function $g_k(\bbH, \bbv^* + \bbepsilon; \bbb^{(k)})$ that we were looking for is given, elementwise, by the second and third terms in the right-hand side of~\eqref{E:proof_necessary_condition_5}. 
	Since $\epsilon_{v_i} \to 0$ as $\eta \to 0$ (equivalently, as $k \to \infty$), it must be that the last term in~\eqref{E:proof_necessary_condition_5} goes to $0$ as $k$ goes to infinity, from where our result follows.	
\end{IEEEproof}

Theorem~\ref{T:convergence_necessary} asserts that if the UWMMSE is uniformly learning the true power allocation $\bbp^*$ as it goes deeper, then the learned weights $\bbb^{(k)}$ must asymptotically satisfy~\eqref{E:convergence_necessary_condition}. 
Intuitively, we expected $\bbb^{(k)} \to \mathbf{0}$ so that the UWMMSE layer boils down to the classical WMMSE iteration. Indeed, $\bbb^{(k)} \to \mathbf{0}$ satisfies the necessary condition in~\eqref{E:convergence_necessary_condition}. Notice, however, that other limits are permitted by this necessary condition, e.g., $b_i^{(k)} \to w_i^*$ for all $i$ would also satisfy~\eqref{E:convergence_necessary_condition}. 
Nonetheless, this would require the function $\Psi$ in~\eqref{E:unfold_1} to be sufficiently expressive to return $\bbw^*$ (an explicit function of the true optimal power allocation) from $\bbH$.
This is not the case in general for the simple function proposed in~\eqref{E:psi}-\eqref{E:psi_2} so that, in practice, we expect~\eqref{E:convergence_necessary_condition} to be satisfied by $\bbb^{(k)} \to \mathbf{0}$.
As is customary in the analysis of deep learning algorithms, the infinite depth architecture is amenable to theoretical analysis but the derived conclusions are also valid in the finite regime. Indeed, as further discussed in Section~\ref{Ss:interpret}, $b_i^{(k)}$ yields values close to zero for most nodes $i$ when $k$ approaches $K$.

Finally, it should be noticed that Theorem~\ref{T:convergence_necessary} is independent of the specific function $\Psi$ chosen in~\eqref{E:unfold_1}.
That is, the above result is a characteristic of the architecture in~\eqref{E:unfold_1}-\eqref{E:unfold_4}, irrespective of the functional form chosen to parameterize $\bbb^{(k)}$.

\subsection{Distributed implementation, scalability, and complexity analysis}\label{Ss:distributed}

It is possible to deploy the trained UWMMSE in a distributed fashion as the allocated power at a given node $i$ can be computed without explicit knowledge of the states of channels that do not include transmitter $i$ or receiver $r(i)$. 
However, similar to WMMSE~\cite{shi2011iteratively}, we need to make some key assumptions regarding the information available to each transceiver pair. 
Primarily, each transmitter $i$ should have estimates of the local channel state $h_{ji}$ for all receivers $r(j)$. 
Additionally, there should be a feedback mechanism through which receiver $r(i)$ can transmit information to all transmitters in the network. 
Under these assumptions, each receiver $r(i)$ can compute its individual $u_i^{(k)}$ and $w_i^{(k)}$ and pass them on to the transmitters to compute $v_i^{(k)}$ for all $k$ unrolling layers in all nodes. 
Each node is also assumed to have access to the full set of trainable weights of both GCN-based learning modules in each unrolling layer $k$. It is important to note here that our proposed method cannot be trained in a distributed fashion as the used formulation of GCN~\cite{kipf2016semi} requires access to the entire CSI matrix and thus necessitates centralized training. But once trained, it is sufficient to have the indexed row and column available for computing the output of any given node, allowing distributed implementation of the trained model. Such a setup would still incur additional communication overheads and the rate of power allocation would be limited by the speed of the feedback link between a given transceiver pair.   

\begin{figure*}[!]
	\centering
	\subfigure[]{
			\centering
			\includegraphics[width=0.31\textwidth,height=0.23\textwidth]{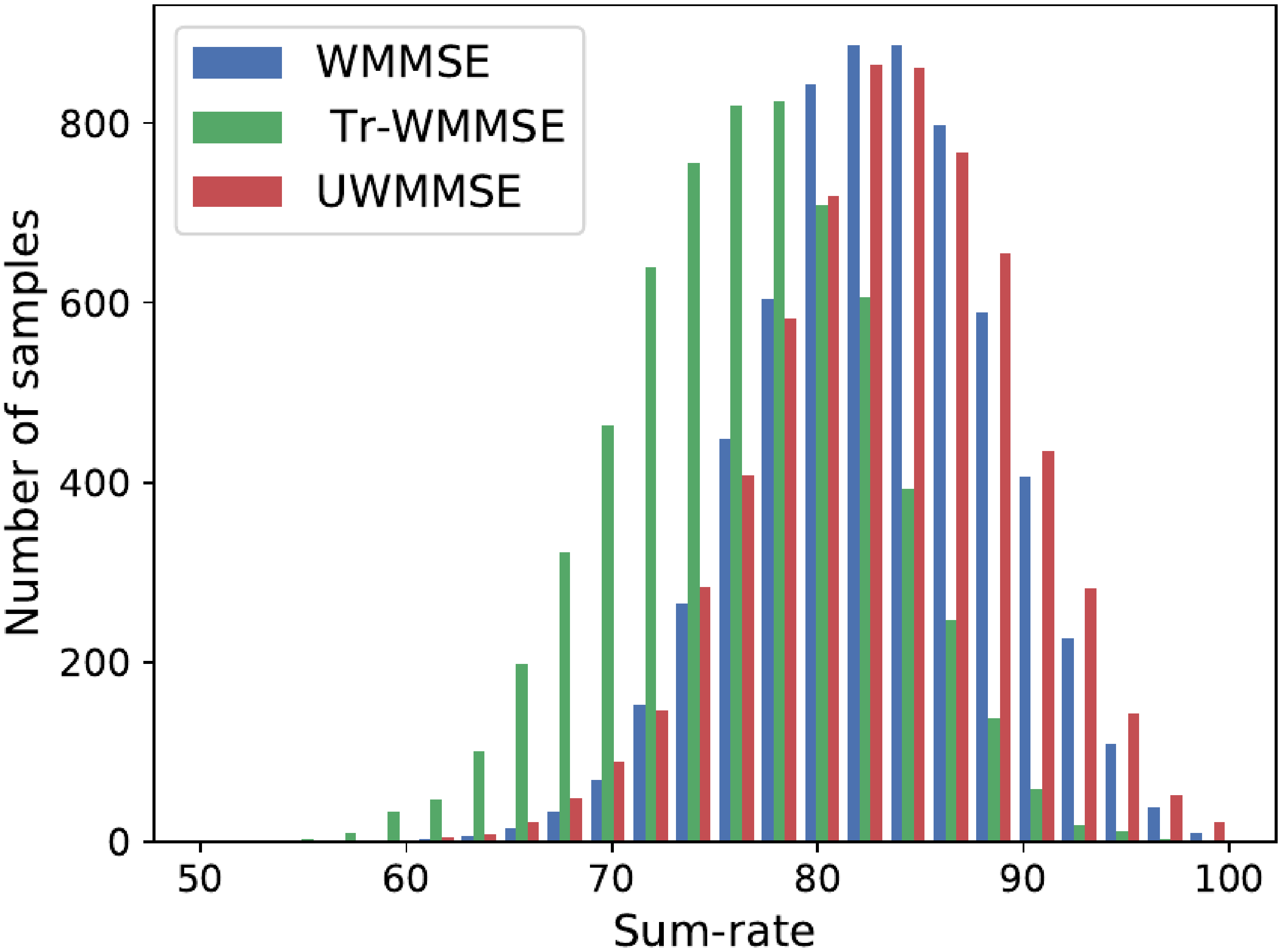}
			\label{Fig:performance_comparison_hist}
		}	
	\subfigure[]{
			\centering
			\includegraphics[width=0.31\textwidth, height=0.23\textwidth]{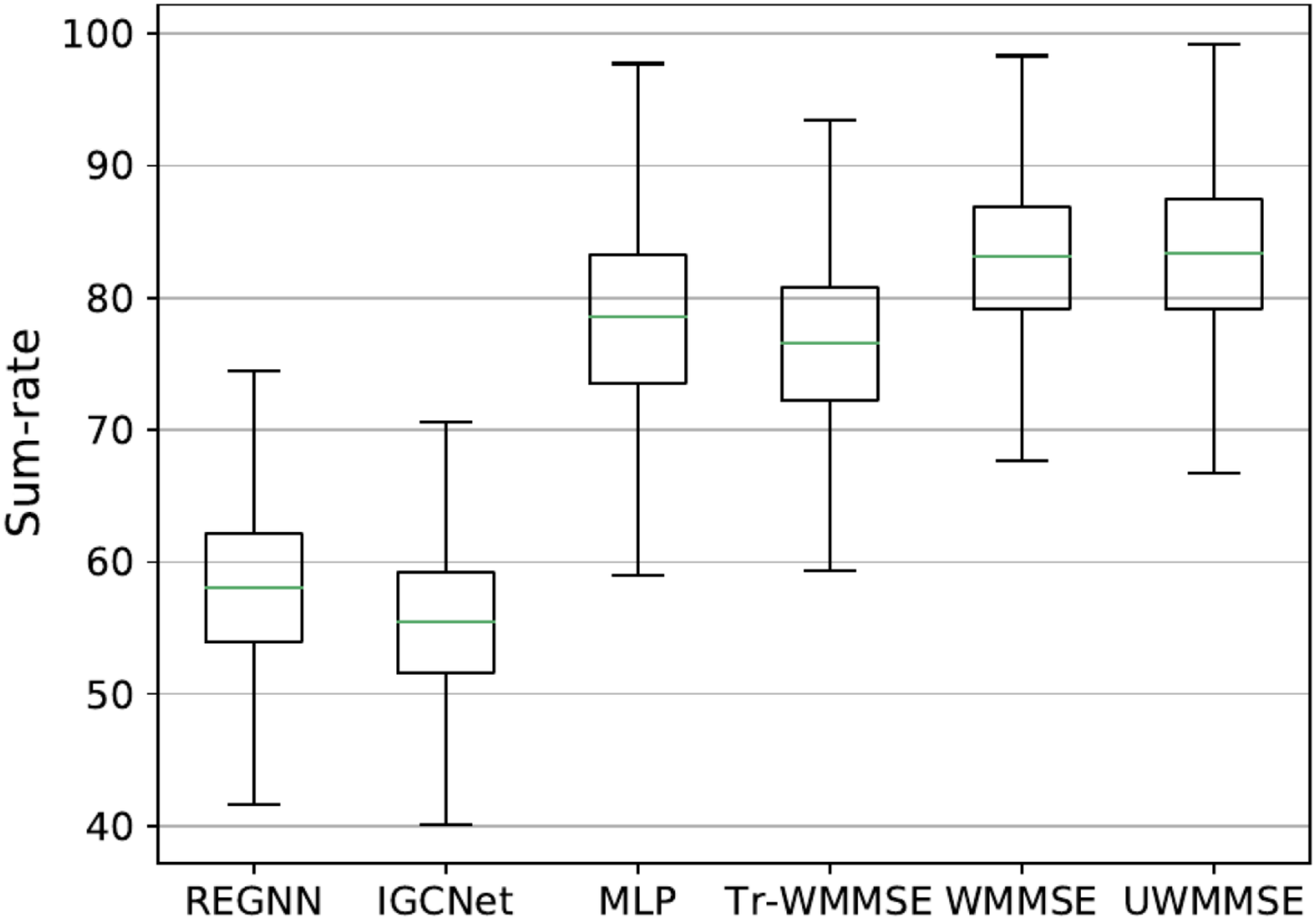}
			\label{Fig:performance_comparison_low_noise}
		}	
	\subfigure[]{
			\centering
			\includegraphics[width=0.30\textwidth, height=0.23\textwidth]{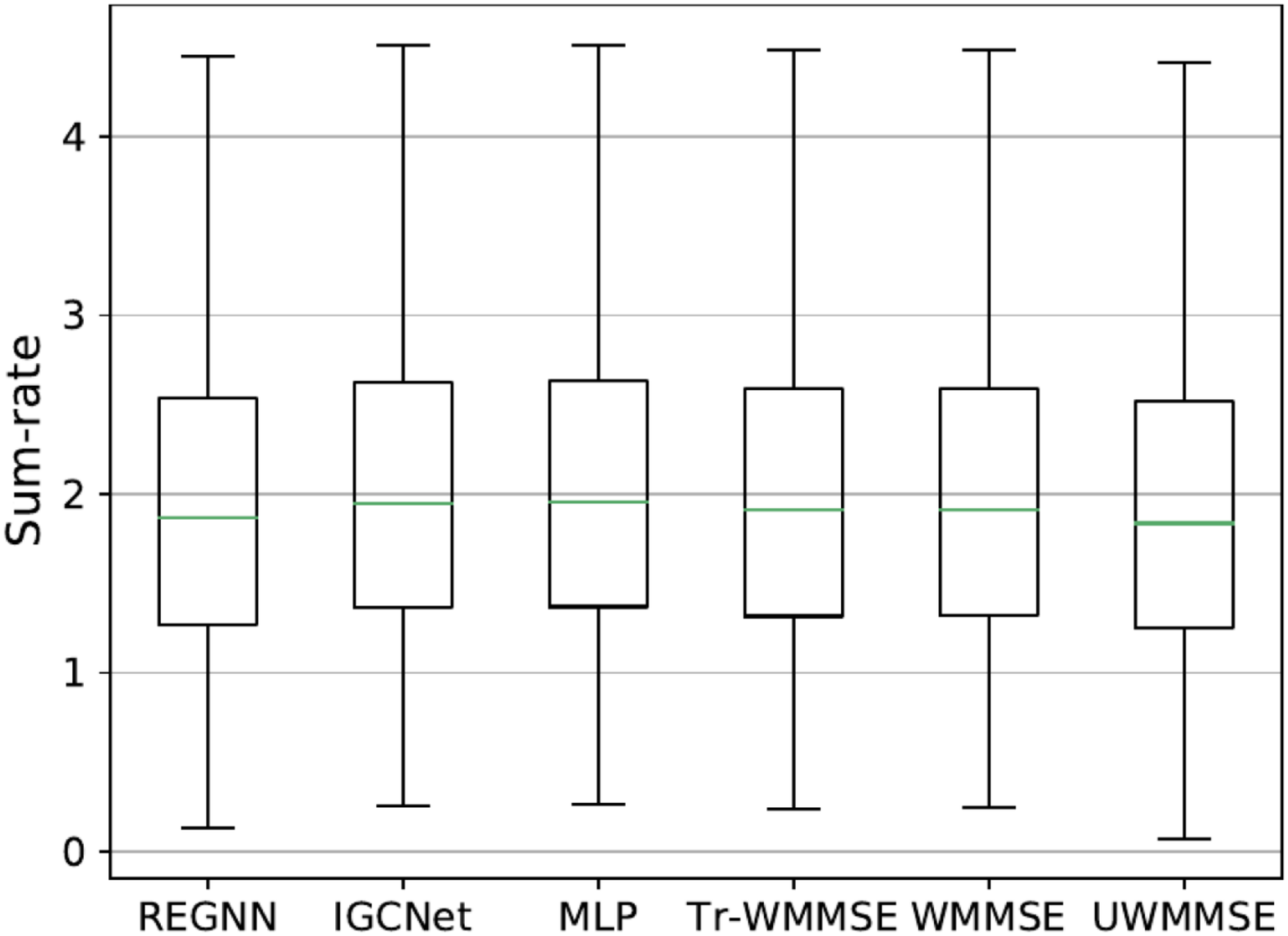}
			\label{Fig:performance_comparison_high_noise}
		}
		\caption{\small {Performance comparison of the proposed UWMMSE with WMMSE \cite{shi2011iteratively}, a truncated version thereof, an MLP \cite{sun2018learning}, IGCNet\cite{shen2019graph}, and REGNN \cite{eisen2020optimal}. 
		(a)~Histogram of the achieved utilities for $6400$ randomly drawn channel state matrices in the low-noise regime ($\sigma =  \num{2.6e-5}$). UWMMSE achieves performance comparable to WMMSE but requiring the same number of iterations than the underperforming truncated version. 
		(b)~Box plots associated with the histograms in (a), where the comparisons with MLP, IGCNet, and REGNN have been added.
		(c)~Counterpart of (b) but for the high-noise regime ($\sigma = 1.0$).}}
		\label{Fig:performance_comparison}
\end{figure*}

The computational complexity per WMMSE iteration has been shown to be $\mathcal{O}(M^2)$ for an $M$-user single-antenna interference network. 
This complexity translates directly to each unrolled layer in UWMMSE as we follow the same update rules as WMMSE. 
In addition, each GCN in a given unrolling layer has a complexity of $\mathcal{O}(M^2F)$ where $F$ is the hidden layer dimension \cite{kipf2016semi}. 
Therefore, the complexity of UWMMSE scales with the size of the network as $\mathcal{O}(M^2)$ per unrolled layer, which is the same as the per-iteration complexity of WMMSE. 
As a result, we achieve a significant gain in terms of processing time by limiting the number of unrolling layers in UWMMSE. This will be validated empirically in Section~\ref{Ss:performance_comparison}.   

\section{Numerical experiments}\label{S:num_exp}

We present extensive numerical experiments to illustrate the performance of UWMMSE in different settings and to validate our understanding of the architecture.\footnote{Code to replicate the numerical experiments here presented can be found at \href{https://github.com/ArCho48/Unrolled-WMMSE.git}{https://github.com/archo48/unrolled-wmmse.git}.}
In Section~\ref{Ss:performance_comparison}, we compare the performance of UWMMSE with competing approaches whereas in Section~\ref{Ss:ablation} we analyze how this performance varies for different choices of key parameters in the proposed architecture.
In Section~\ref{Ss:node_feat_utility}, we demonstrate how the considered framework can be enriched by incorporating node features and varying the utility function.
Our theoretical result in Theorem~\ref{T:convergence_necessary} is empirically validated in Section~\ref{Ss:interpret}. 
Finally, in Section~\ref{Ss:spat_den} we analyze the generalizability of the proposed approach to changes in the density and size of the wireless network under consideration.

For all our experiments, we consider a \textit{Rayleigh fading} channel model. 
To that end, we construct a random geometric graph in two dimensions having $M$ transceiver pairs. 
First, each transmitter $i$ is dropped uniformly at random at locations $\mathbf{t}_i \in [-M, M]^2$. 
Then, its paired receiver $r(i)$ is dropped uniformly at random at location $\mathbf{r}_i \in [\bbt_i - \frac{M}{4}, \bbt_i + \frac{M}{4}]^2$. 
Under fading conditions, the channel between a transmitter $i$ and any receiver $r(j)$ is composed of two components $H_{ij} = H_{ij}^P H_{ij}^F$, where the path gain is given by $H_{ij}^P = \lVert \mathbf{t}_i - \mathbf{r}_j \rVert ^ {-2.2}$ and the fading coefficient is randomly drawn $H_{ij}^F \sim \text{Rayleigh}(1)$. 
We fix $H_{ij}^P$ and sample $H_{ij}^F$ at each power allocation instance. 
In all but the experiments presented in Section~\ref{Ss:spat_den}, we assume that the underlying topology of the network is fixed. 
This can be interpreted as a set of static nodes communicating with each other over a period of time. 
At each discrete-time instant, fading conditions change and the instantaneous channel information is used by the model to generate the optimal power allocation. 
In all but the experiment presented in Section~\ref{Ss:node_feat_utility}, we set the node features to unity (cf. Remark~\ref{R:node_features}).

Our proposed solution model, UWMMSE, is composed of $4$ unrolled WMMSE layers with each layer $k$ having two $2$-layered GCNs~\eqref{E:psi}-\eqref{E:psi_2}  as its learning modules for the affine transform parameters $\bba^{(k)}$ and $\bbb^{(k)}$. 
Hidden layer dimension of all GCNs is set to $4$. 
ADAM~\cite{kingma2014adam} optimizer is employed for training across $10000$ steps with each training step being performed on a batch of $64$ randomly sampled channel fading states. 
Learning rate is set to \num{1e-3} and training is performed for a maximum of $20$ epochs, with early stopping. 
For testing, we randomly sample $6400$ channel fading states. Unless otherwise specified, the network size $M$ is fixed at $20$.

\subsection{Performance comparison}\label{Ss:performance_comparison}

We begin by comparing the performance attained by UWMMSE with that of established baselines and state-of-the-art methods.
We perform these comparisons in two different regimes that are defined on the basis of additive channel noise variance. 
In the \textit{high-noise} regime, the standard deviation ${\sigma}$ of the noise [cf.~\eqref{E:data_rate}] is set as ${1}$ whereas in the \textit{low-noise} regime $\sigma = \num{2.6e-5}$. 

We choose the following methods for comparison, all of which address the problem of optimal power allocation in wireless networks with minor variations in noise regimes and channel models, achieving state-of-the-art performance on standard benchmarks.

\begin{figure*}
	\centering
	\subfigure[]{
			\centering
			\includegraphics[width=0.30\textwidth]{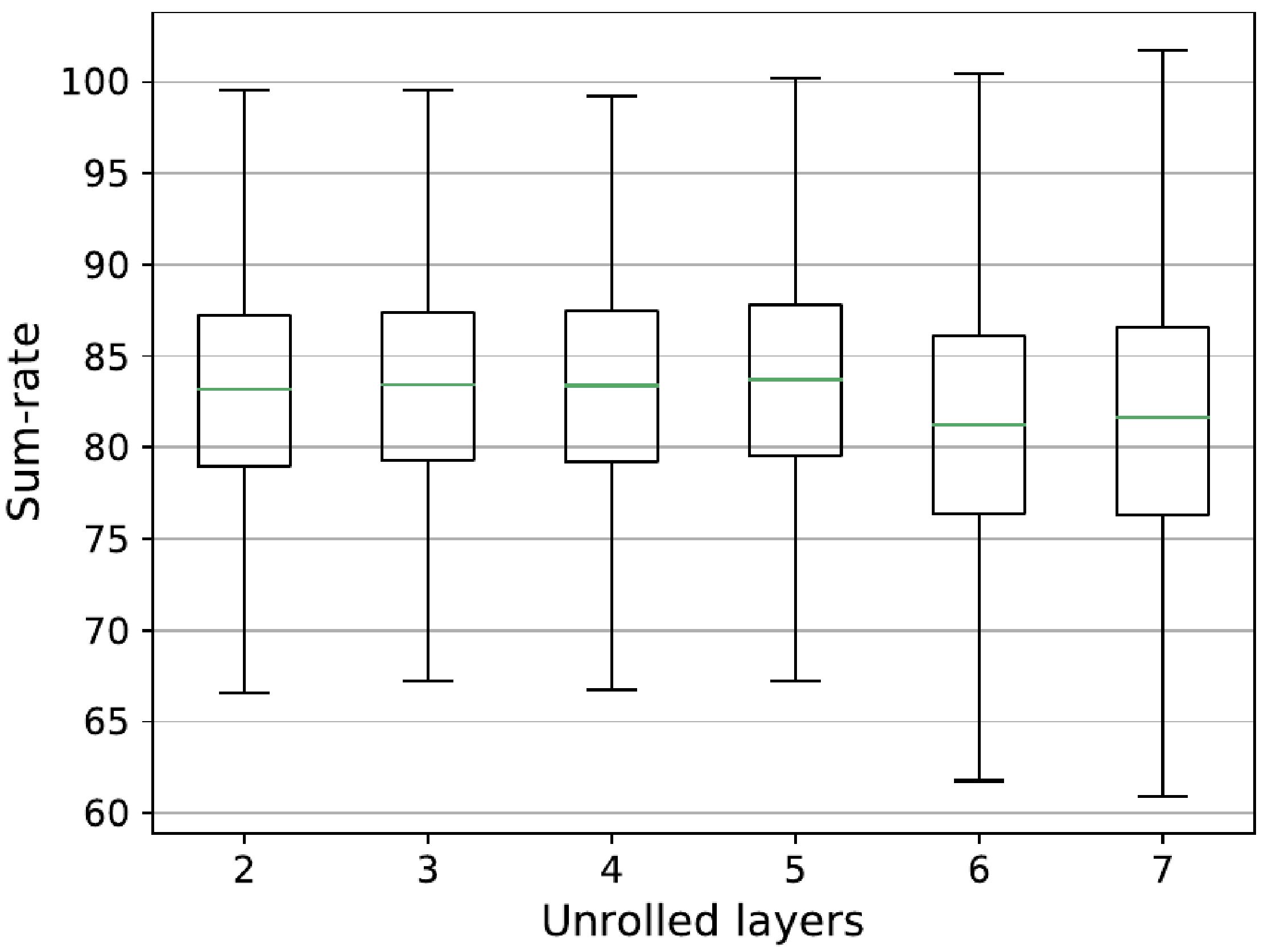}
			\label{Fig:depth}
		}	
	\subfigure[]{
			\centering
			\includegraphics[width=0.30\textwidth]{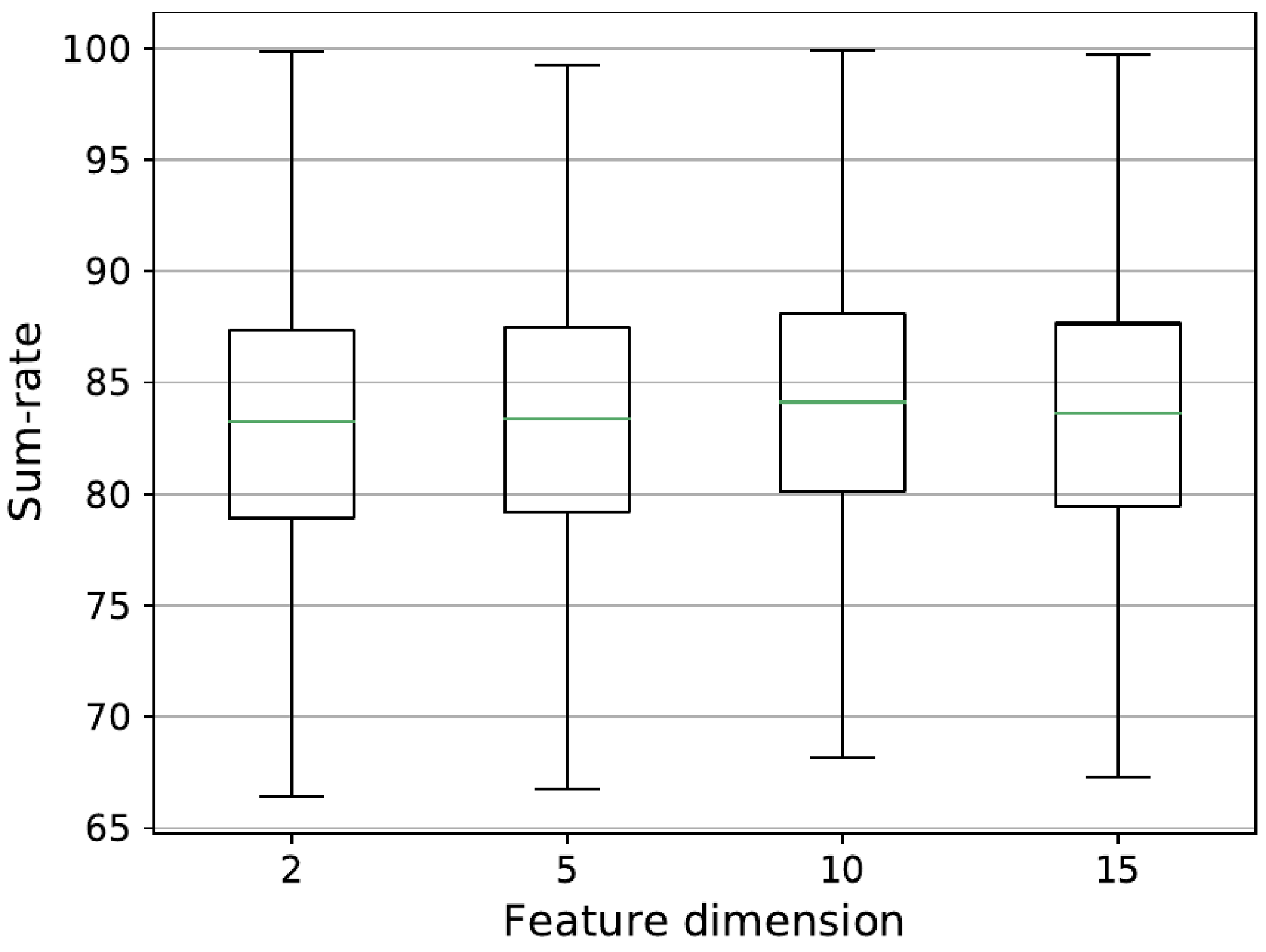}
			\label{Fig:width}
		}	
	\subfigure[]{
			\centering
			\includegraphics[width=0.31\textwidth]{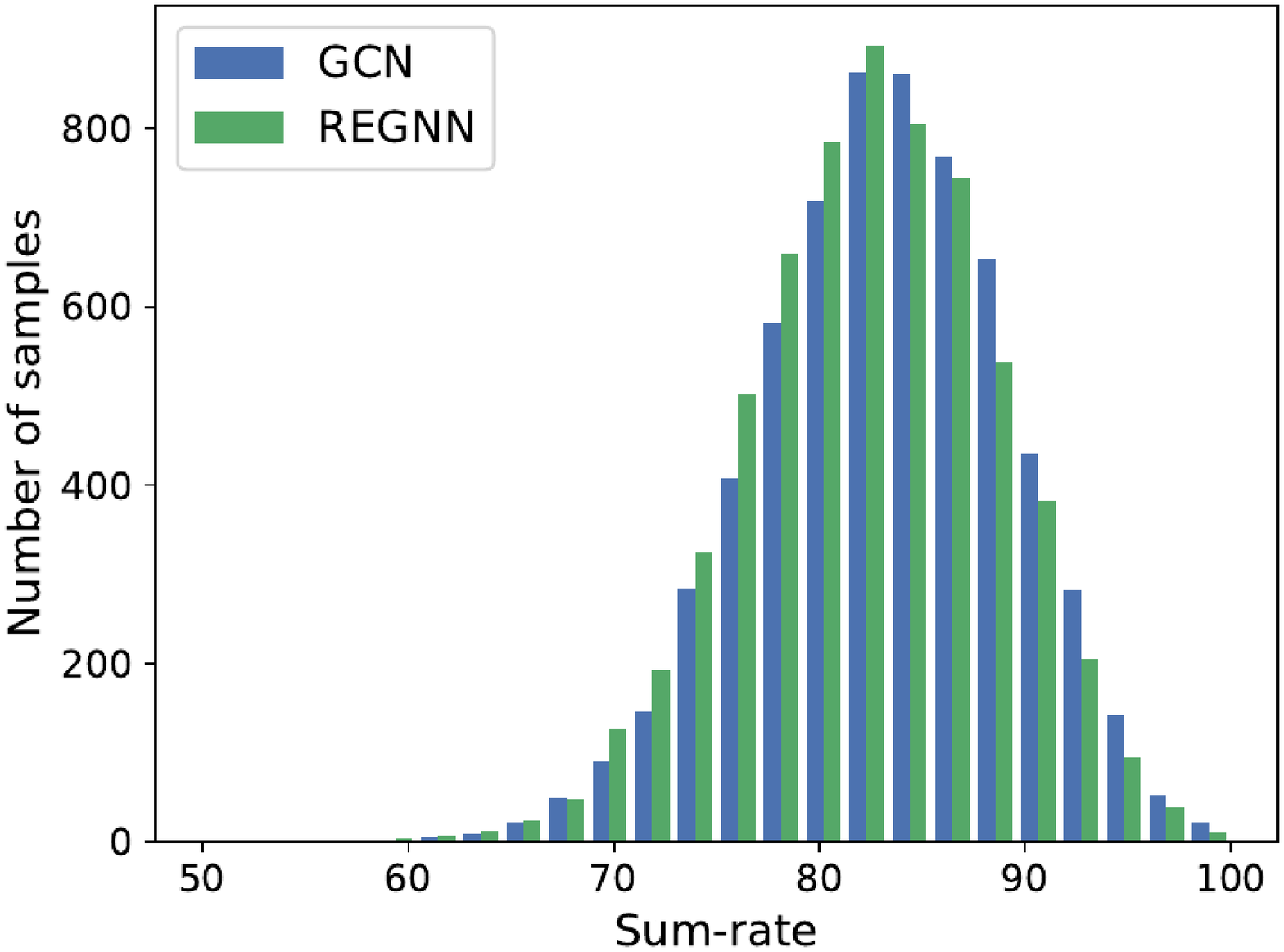}
			\label{Fig:gnn}
		}
		\caption{\small Impact of variation in \textit{depth}, \textit{width} and \textit{GNN architecture} on UWMMSE performance. 
	    (a)~Box plots of sum-rate achieved by UWMMSE with number of unrolled layers varying from $2$ to $7$.
	    (b)~Box plots of sum-rate achieved by UWMMSE with the dimension of GCN hidden features varying from $2$ to $15$.
	    (c)~Histogram of sum-rate achieved by UWMMSE with a GCN-based learning module versus that with an REGNN-based learning module.
		}
		\label{Fig:impact}
\end{figure*}

\begin{enumerate}
    \item \textit{WMMSE} \cite{shi2011iteratively} forms the baseline for our experiments as it is the most commonly followed classical approach to address this problem. We set a maximum of $100$ iterations for WMMSE per sample.
    \item \textit{Truncated WMMSE} (Tr-WMMSE) provides a performance lower bound to UWMMSE in terms of the  sum-capacity that WMMSE achieves with as many iterations as the unrolled layers without the learning component. We fix the number of iterations of Tr-WMMSE to $4$ to match UWMMSE unrollings.   
    \item \textit{MLP} \cite{sun2018learning} is employed in a supervised setting to learn a functional mapping between channel state information and corresponding power allocation, using WMMSE output as the training signal.
    \item \textit{REGNN} \cite{eisen2020optimal} addresses the specific problem of \textit{binary} power allocation in Rayleigh interference channels. It tackles this problem in the high-noise regime where it achieves state-of-the-art performance. 
    \item \textit{IGCNet} \cite{shen2019graph} addresses the power allocation problem in Gaussian interference channels, specifically, in the high-noise regime, where it achieves state-of-the-art performance.  

\end{enumerate}

The comparisons are shown in Fig~\ref{Fig:performance_comparison}. 
Owing to the randomness of channel state information, there can be significant variations in the utility function value across test samples for any given power allocation algorithm. 
Each individual channel realization can be more or less challenging based on how the interference strengths compare with the channel strengths for the specific fading conditions. 
To illustrate this, we show a histogram of utility values over the entire test set in the low-noise regime; see Fig~\ref{Fig:performance_comparison_hist}. 
The empirical distribution of utilities across different channel instances shows that UWMMSE matches the performance of WMMSE for most realizations, while even surpassing it in a few cases. 
To be more precise, UWMMSE, WMMSE, and Tr-WMMSE respectively achieve average sum-rates of $83.21$, $82.94$, and $76.49$ as shown in Table~\ref{tab:performance} with corresponding standard deviations of $6.10$, $5.76$, and $6.28$.
The performance gain in WMMSE compared to its truncated version was expected due to additional iterations that add to the complexity of the solution. 
UWMMSE bridges this gap with a learning module that performs more intelligent update steps to converge to the optimal within as many iterations as the underperforming Tr-WMMSE.

Having compared UWMMSE with the classical method, we now expand the analysis to include other learning-based approaches in the low-noise regime; see Fig~\ref{Fig:performance_comparison_low_noise}. 
MLP, which learns a functional mapping of the WMMSE output, beats Tr-WMMSE with an average sum-rate of $78.17$, still falling short of WMMSE by a significant margin. 
This shows that supervised methods are limited by the quality of their training signals and often do not generalize to out-of-sample data. 
On the other hand, both REGNN and IGCNet -- originally designed and tested in high-noise regimes -- prove to be inadequate to match the performance of WMMSE in the more challenging low-noise setting. 
For instance, REGNN was designed for binary power allocation -- either $0$ or $p_{\max}$ -- but finer precision is needed for state-of-the-art performance in the low-noise regime.
Indeed, in this regime the interference component dominates in the sum-capacity formulation [cf.~\eqref{E:data_rate}] and, therefore, minor variations in power allocation result in significant performance alteration.  

By contrast, we now shift focus to the high-noise regime, where noise dominates over the interference component in the SINR formulation. Hence, the sum-capacity achieved is less dependent on the precise power allocation at each node. 
In this setting, all the algorithms tend to follow a binary approach in which each node would either transmit with full power or not transmit at all. 
Due to the high noise in this setting, low performance is generally achieved as clearly manifested by the low sum-rate values obtained by all the algorithms in Fig~\ref{Fig:performance_comparison_high_noise}. 
Also, given the high noise, it is always optimal to transmit with almost full power at most nodes. Consequently, there is little performance variation across methods. 

\begin{figure*}
	\centering
	\subfigure[]{
			\centering
			\includegraphics[width=0.30\textwidth,height=0.23\textwidth]{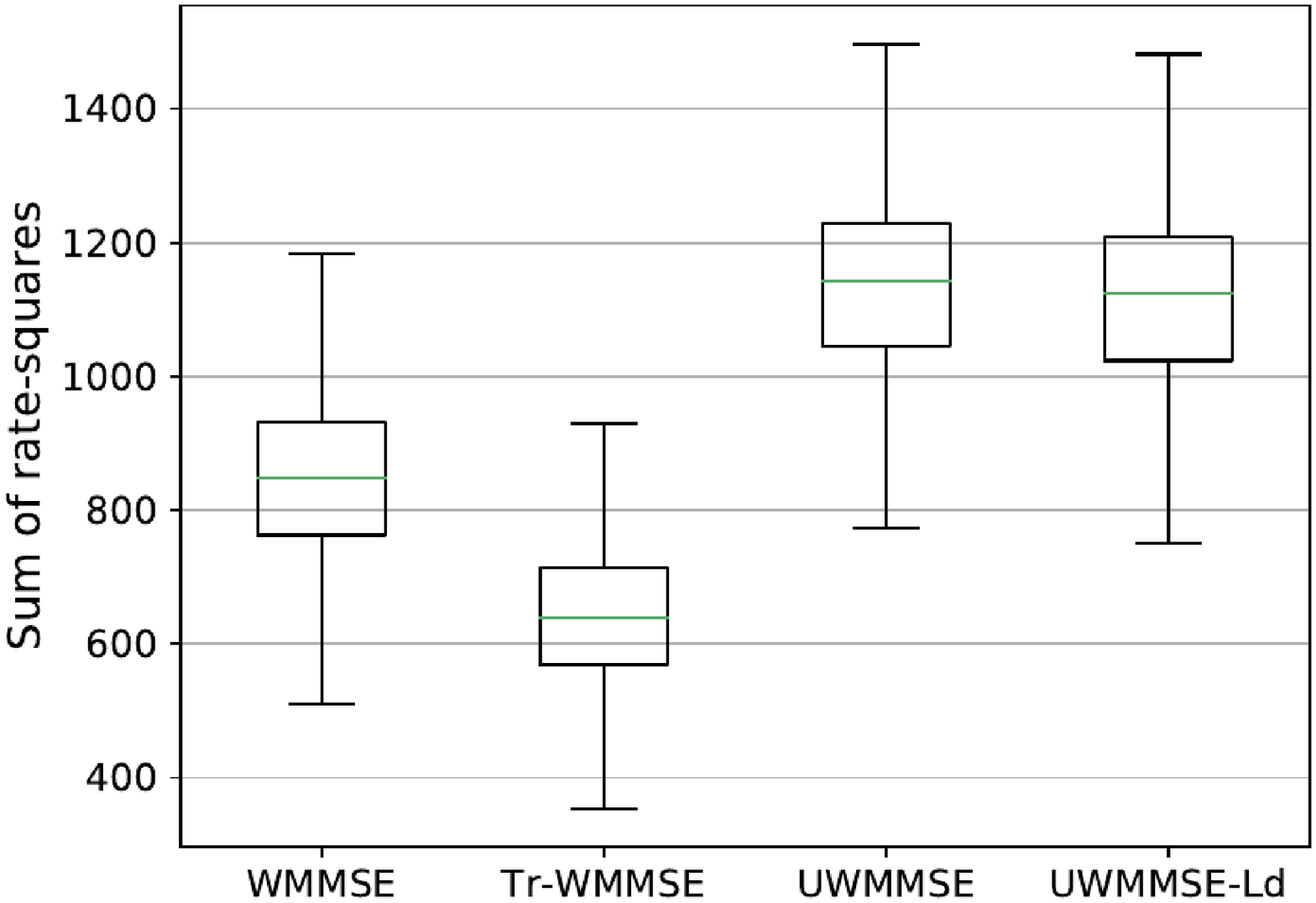}
			\label{Fig:utility1}
		}
	\subfigure[]{
			\centering
			\includegraphics[width=0.30\textwidth,height=0.23\textwidth]{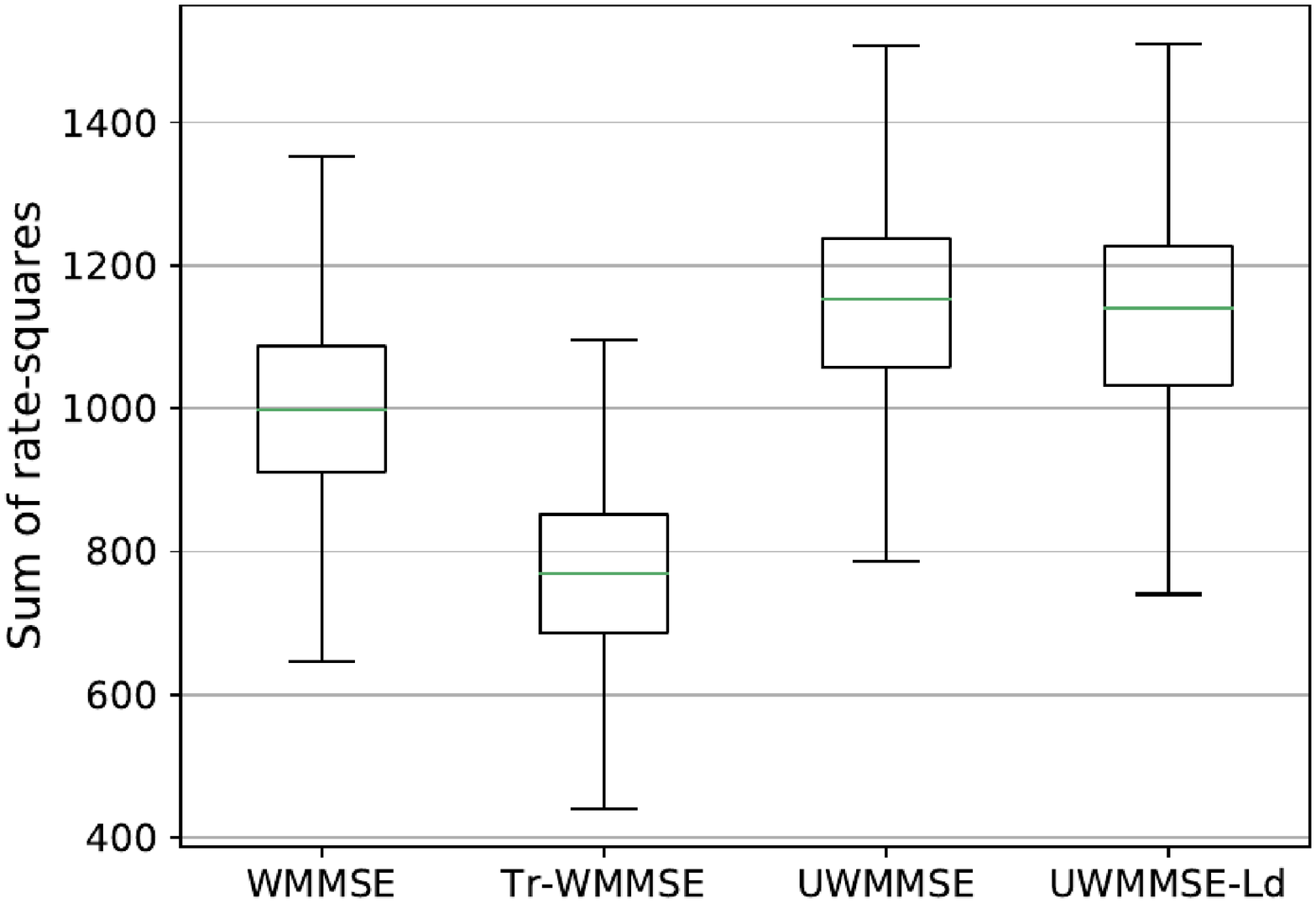}
			\label{Fig:utility2}
		}
	\subfigure[]{
			\centering
			\includegraphics[width=0.30\textwidth,height=0.23\textwidth]{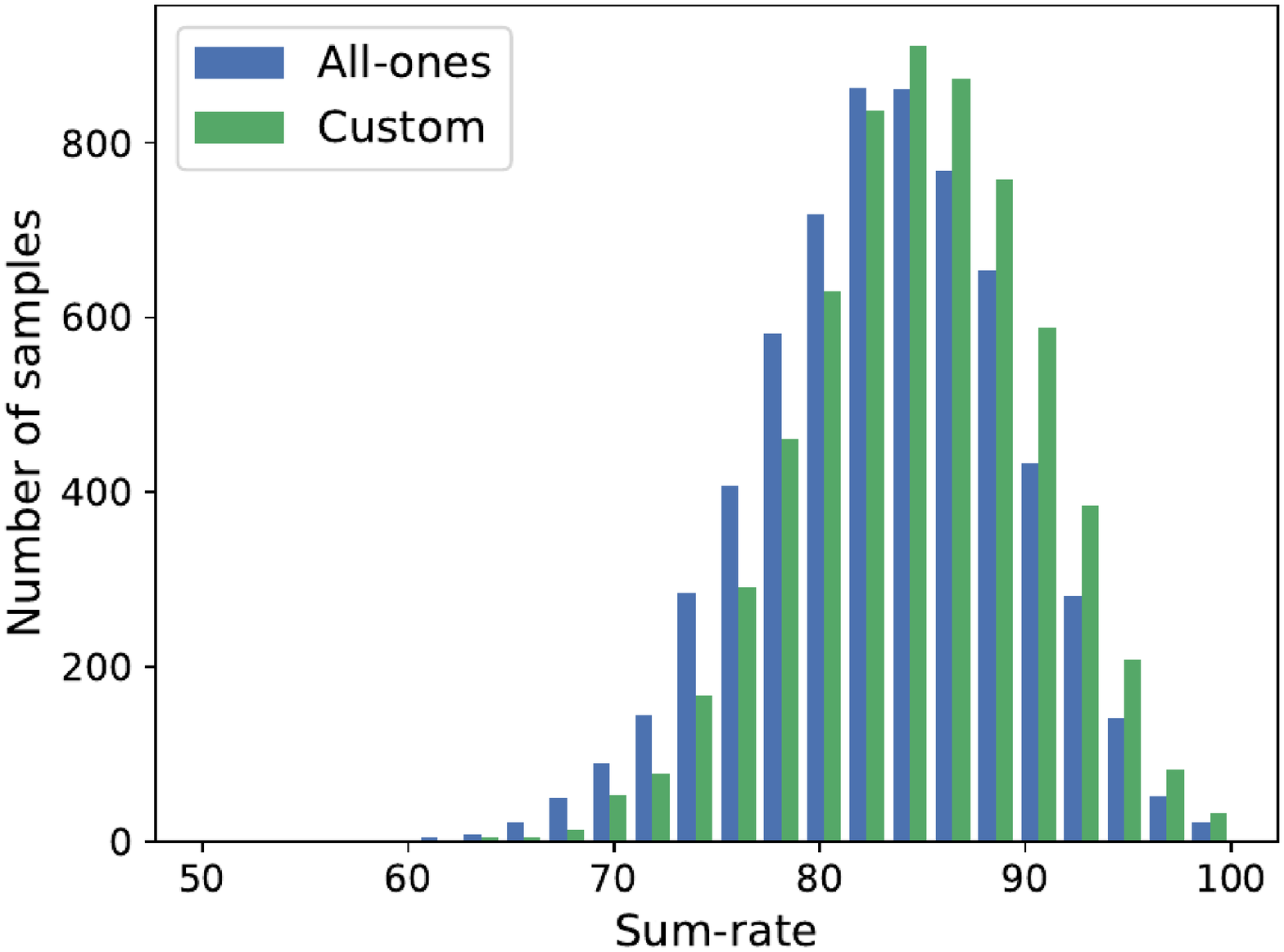}
			\label{Fig:nodefeat}
		}
		\caption{\small Experimental validation of Remarks~\ref{R:different_utilities} and~\ref{R:node_features}.  
		(a)~Boxplot comparing the performance of UWMMSE, WMMSE, Tr-WMMSE, and a version of UWMMSE trained with less data (UWMMSE-Ld), in terms of the sum of squared rates, without the proposed modification in~\eqref{E:generic_w} for the update of $\bbw^{(k)}$. 
		(b)~Counterpart of (a) when the proposed modification in~\eqref{E:generic_w} is used.
		(c)~Histogram of UWMMSE performance with all-one node features $\bbQ = \mathbf{1}$ versus that with customized node features. 
		}
		\label{Fig:incorporating}
\end{figure*}

\begin{table}[!htbp]
\centering
\caption{Performance and time comparisons of all methods}
\begin{tabular}{|l | c | c | c | c | c|}
\hline
Algorithm  & Training  & Test & Test \\
& time (min) & sum-rate & time (msec) \\
\hline \hline
WMMSE~\cite{shi2011iteratively} &  -  &  82.94  & 16 \\ 
Tr-WMMSE &  -  &   76.49  & 1.0 \\
MLP~\cite{sun2018learning} & 0.5 &  78.17   &  3.2 \\
REGNN~\cite{eisen2020optimal} &  15  &  57.92   & 2.5  \\
IGCNet~\cite{shen2019graph} &  5   &  55.30   & 3.0  \\
UWMMSE & 15 &  \textbf{83.21}  & \textbf{2.0} \\
\hline
\end{tabular}
\label{tab:performance}
\end{table}

Achieving a sum-rate that is close to optimal is not enough as the time required for power allocation has to be synchronized with fading conditions that are fast evolving. It is therefore essential to consider the time taken by each algorithm to offer a power allocation output given a channel state input. 
To that end, we provide a computation time comparison\footnote{All computations were performed on an Nvidia Quadro T2000 GPU.} among the various algorithms in Table~\ref{tab:performance}.  UWMMSE, which takes close to $2$ ms per sample, is significantly faster as compared to WMMSE which takes around $16$ ms per sample. 
All the other learning-based methods~ \cite{shen2019graph,sun2018learning,eisen2020optimal} have a processing time similar to that of UWMMSE, however, none of them achieve the same performance in that time duration, which is the main advantage of our method over existing algorithms. Unlike WMMSE and similar to any other learning-based method, UWMMSE has the additional requirement of training, which takes close to $15$ mins, but in most applications this is not a problem since we envision the  computational effort to be done prior to implementation. MLP~\cite{sun2018learning} has significantly lower training time, but requires a full iteration of WMMSE over its training data to get the ground-truth power allocation which adds to the time complexity.  

\subsection{Impact of depth, width, and GNN functional form}\label{Ss:ablation}

Our next objective is to analyze the dependence of our proposed method on the various design parameters in the architecture. To that end, we study the variation of UWMMSE performance specifically with respect to two architectural components, namely the number of unrolled layers (\textit{depth}) and the dimension of hidden features (\textit{width}). 
We evaluate UWMMSE performance separately for each of these hyper-parameters taking a fixed set of values, $depth \in \{2,3,4,5,6,7\}$ and $width \in \{2,5,10,15\}$, in an experimental setup that is otherwise identical to that in Section~\ref{Ss:performance_comparison}. 
As shown in Fig~\ref{Fig:depth}, mean sum-rate improves from $2$-layered to $5$-layered architectures and then decays as more layers are added to the model. 
This could be attributed to the need of more data to successfully train deeper architectures.
On the other hand, in Fig~\ref{Fig:width} we observe overlapping distributions of sum-rate for all the configurations with minor improvement on using a hidden feature dimension of $10$. 
The observed robustness in terms of depth and width of the architecture is of practical importance, providing the practitioner with a wide design space without compromising performance.


We further evaluate the performance of our proposed unrolling scheme by employing an alternative learning module, i.e., selecting a different function $\Psi$ in~\eqref{E:unfold_1}. 
Our goal is to characterize how the achieved performance depends on the specific selection of the GNN.
To that end, we compare our canonical choice \eqref{E:psi}-\eqref{E:psi_2} with an REGNN~\cite{eisen2020optimal}, primarily due to its simple and lightweight architecture.
REGNN also interprets the CSI as the adjacency matrix of a graph and constructs a layered architecture by intertwining linear graph filters with point-wise non-linearities. 
To be more precise, if we denote by $\bbz_l$ the output of an intermediate layer in an REGNN, we have that
\begin{equation}\label{E:REGNN}
    \bbz_l = \alpha\left( \sum_{k=0}^K \nu_{lk} \bbH^k \bbz_{l-1}\right),
\end{equation}
where $\alpha$ is a ReLU activation function and the coefficients $\nu_{lk}$ are trained. The REGNN is defined by recursive application of~\eqref{E:REGNN} and we test it here as an alternative to GCNs for the computation of $\bba^{(k)}$ and $\bbb^{(k)}$ in~\eqref{E:unfold_1}.

As shown in Fig.~\ref{Fig:gnn}, the empirical distribution of sum-rate achieved by the REGNN-based UWMMSE model has significant overlap with that of the GCN-based baseline. 
These neural networks can therefore be used interchangeably to achieve very similar performances revealing that the proposed unfolded architecture is not restricted by a specific choice of GNN. 
In this way, the practitioner can choose the GNN of preference given the application-specific computation and memory constraints.
It is important to notice that once the REGNN is incorporated into our unfolded architecture, the performance markedly increases to its standalone implementation [cf. Figs.~\ref{Fig:performance_comparison_low_noise} and~\ref{Fig:gnn}] unveiling the importance of including domain-specific knowledge in the neural architecture.

\begin{figure*}
	\centering
	\subfigure[]{
			\centering
			\includegraphics[width=0.30\textwidth]{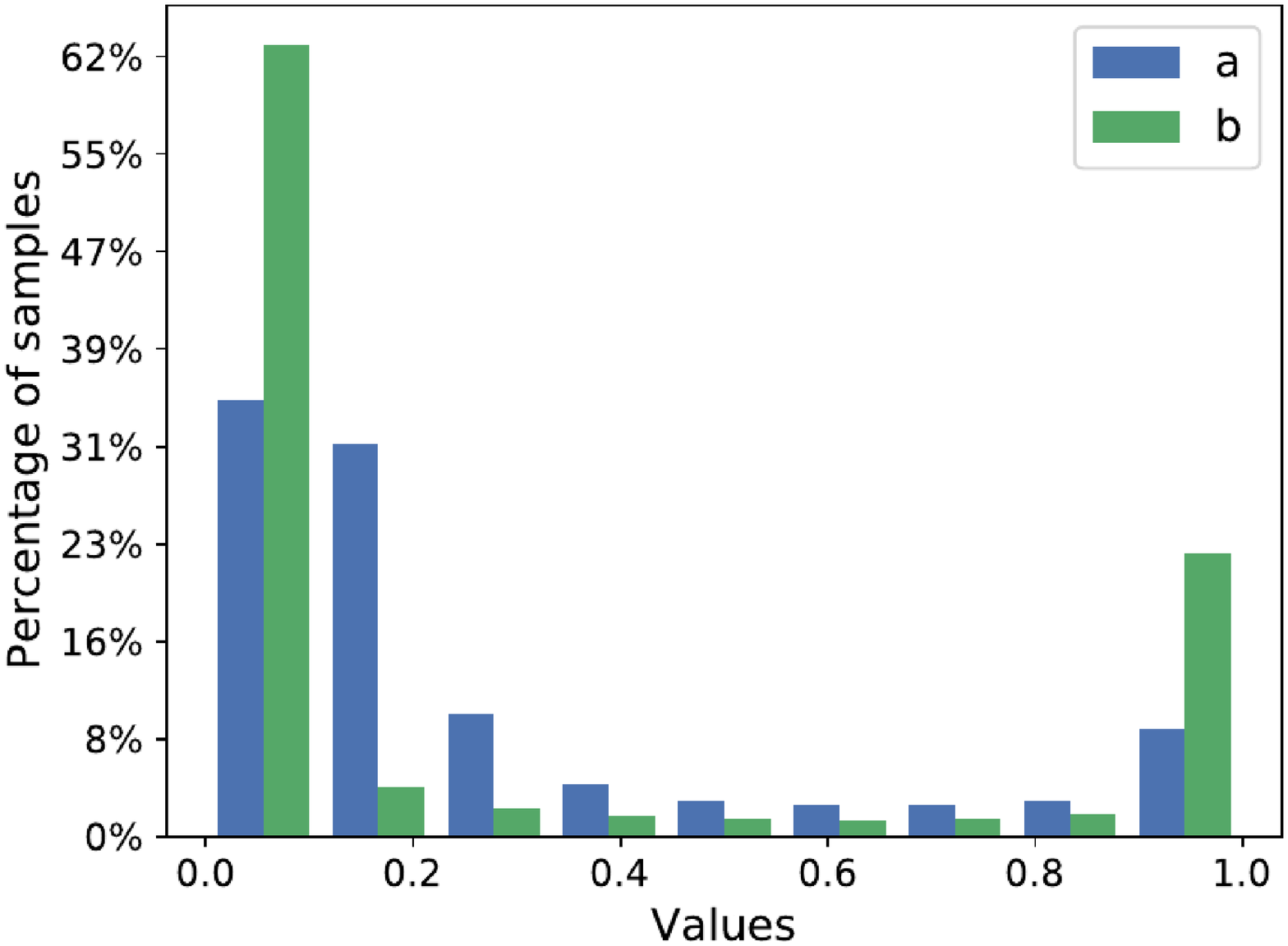}
			\label{Fig:l1}
		}	
	\subfigure[]{
			\centering
			\includegraphics[width=0.30\textwidth]{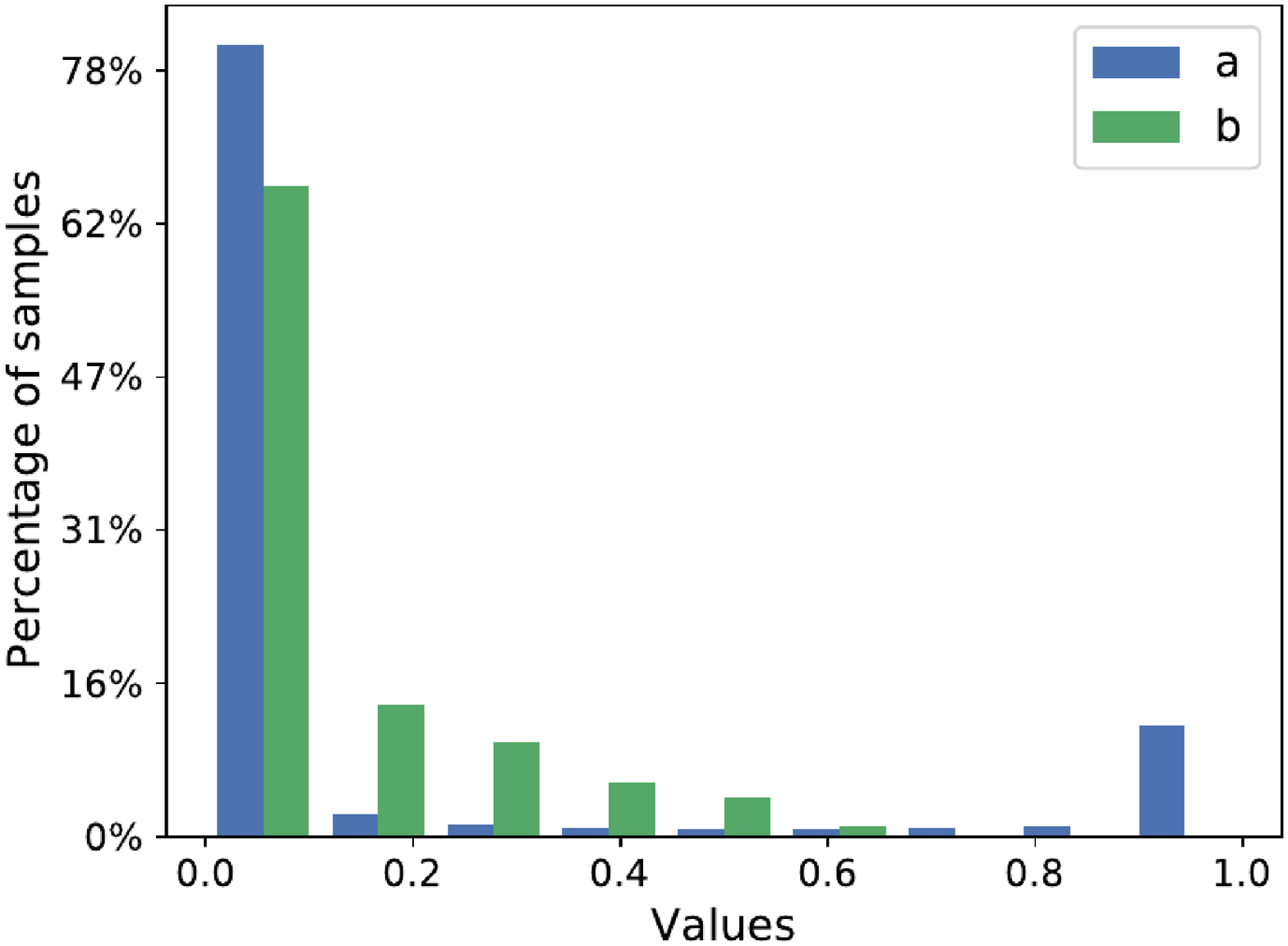}
			\label{Fig:l3}
		}
	\subfigure[]{
			\centering
			\includegraphics[width=0.30\textwidth]{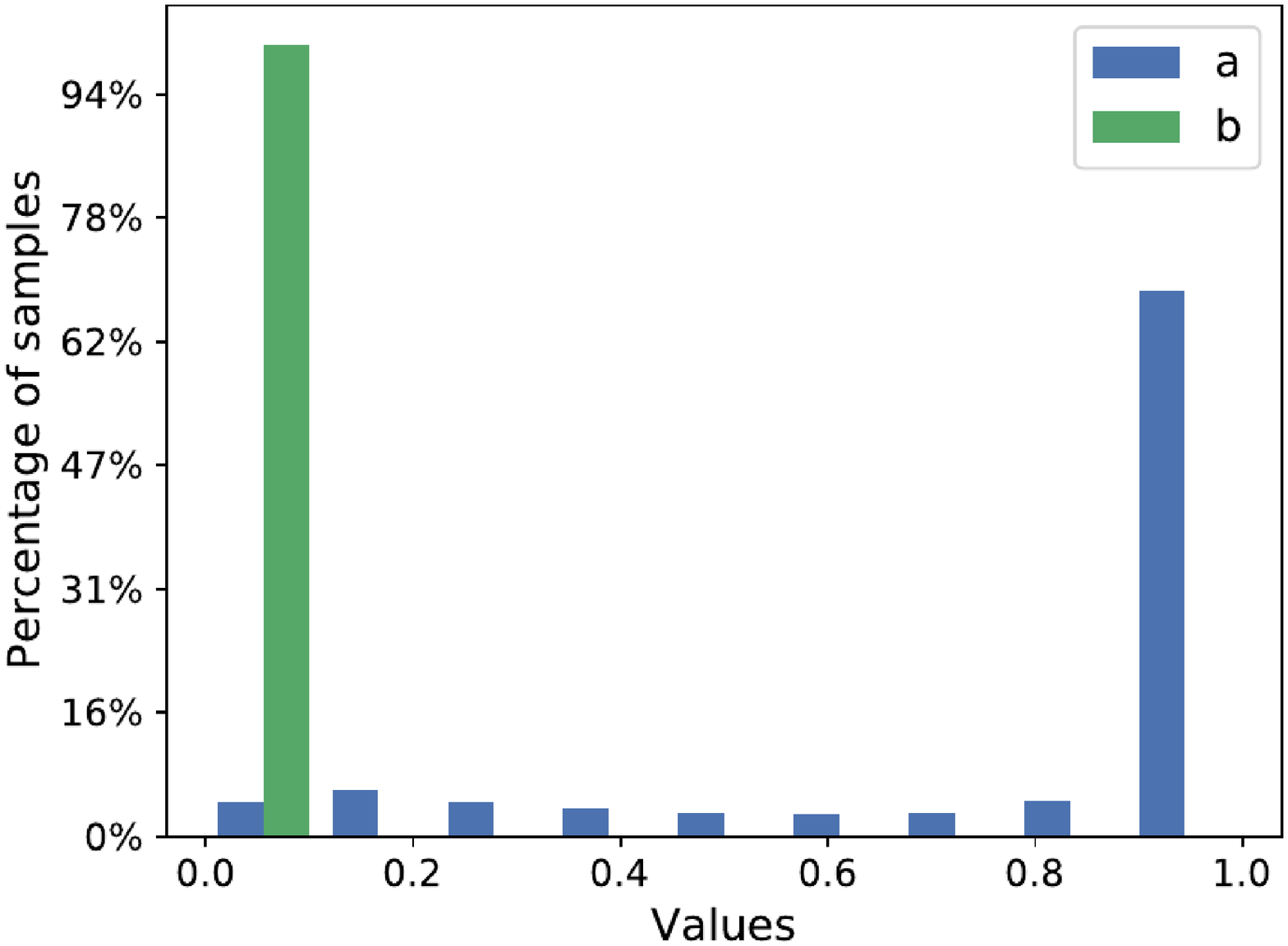}
			\label{Fig:l5}
		}
		\caption{\small Histograms of learned variables $\bba^{(k)}$ and $\bbb^{(k)}$ in layers $k \in \{1,3,5\}$ of a fully trained $5$-layered UWMMSE model.
		(a)~In layer-$1$, the majority of $\bba^{(1)}$ and $\bbb^{(1)}$ values are concentrated at the lower end while almost one-third of the $\bbb$ values are distributed at the higher end of the range.
		(b)~Layer-$3$ sees a transition of $\bbb^{(3)}$ values towards the lower end of the range.
		(c)~Layer-$5$ presents a clear separation in which the majority of $\bba^{(5)}$ values are concentrated near $1$ while almost all $\bbb^{(5)}$ values move to $0$, marking the convergence of UWMMSE to the original WMMSE updates.
		}
		\label{Fig:empirical}
\end{figure*}

\subsection{Incorporating node features and different utility functions}\label{Ss:node_feat_utility} 

We discuss in Remark~\ref{R:different_utilities} the extension of our unrolled scheme to several other utility functions $\beta(z)$ provided that they satisfy the condition that $\gamma(z)$ is strictly concave for $z>0$, where $\gamma(z) = -\beta(-\log(z))$. 
To validate this, we select the sum of squares of data rate as our utility function $\beta(z) = z^2$. 
In this case, $\gamma(z) = -(\log(z))^2$ is strictly concave for $z>0$, as required. 
Under the same experimental setup as in Section~\ref{Ss:performance_comparison}, our model now learns to maximize the sum of squared rates as its utility function. 
From comparing Figs.~\ref{Fig:utility1} and~\ref{Fig:utility2}, one can observe a clear improvement in performance of WMMSE and its truncated version when the modification proposed in~\eqref{E:generic_w} is employed, as expected.
More interestingly, UWMMSE can attain high performance even when the modified update is not used [cf. Fig.~\ref{Fig:utility1}].
This points towards the fact that the learnable parameters $\bba^{(k)}, \bbb^{(k)}$ can rely on data to enforce the right update trajectory for $\bbw^{(k)}$ even in the presence of a model mismatch.
Indeed, the improvement of UWMMSE with the modified update compared to the version without it is minimal (means of $1141.38$ and $1131.86$, respectively).
This difference is slightly magnified in the presence of less training data, where the ability of the learning parameters to make up for the model mismatch is restricted.
To illustrate this, we employed a variant of UWMMSE which was trained on half of the available data (UWMMSE-Ld). 
As expected, UWMMSE-Ld underperforms with respect to UWMMSE in both settings. When comparing the rates for UWMMSE-Ld with and without the modified update, we observe that the modified update yields a utility of 1124.58 whereas without the update the utility is of 1112.42.
It should be noticed that there is still a major improvement with respect to the classical WMMSE, underscoring the importance of the learning module even with limited training data.

Remark~\ref{R:node_features} analyzes the possibility of enriching UWMMSE with node attributes that may encode essential information pertaining to a specific node such as priority or queue length. 
To validate the effectiveness of such a formulation, we perform a simple experiment in which $\bbQ = \mathbf{1}$ (cf. Remark~\ref{R:node_features}) is augmented with two custom features for each node, namely, physical distance between a transmitter and its corresponding receiver ($\lVert \bbt_i - \bbr_i \rVert$) and the shortest distance between each transmitter and all other receivers ($\min_{j \neq i} \lVert \bbt_i - \bbr_j \rVert$). 
Intuitively, a transmitter's average power allocation depends on the distance to its intended receiver as well as the distance to other receivers, since the latter determines the interference effect. Hence, we expect the learning method to take advantage of having explicit access to these features, as illustrated in Fig.~\ref{Fig:nodefeat}. 
Indeed, the mean sum-rate achieved using the features is of $84.73$ compared to the baseline utility of $83.21$.
This result further emphasizes the benefits of infusing domain knowledge in learning-based solutions, where the specific choice of features relies on our understanding of the problem at hand.

\subsection{Empirical validation of Theorem~\ref{T:convergence_necessary}}\label{Ss:interpret}

To develop a deeper understanding of the working dynamics of our proposed approach, we observe the values that the learned variables take in each unrolled layer. 
In this process, we also develop an empirical validation of Theorem~\ref{T:convergence_necessary}. To that end, we retrieve $\bba^{(k)}$ and $\bbb^{(k)}$ values for layers $k \in \{1,3,5\}$ of a fully trained $5$-layered UWMMSE model and plot the corresponding histograms in Fig~\ref{Fig:empirical}. 
Each histogram is built based on $12800$ values, corresponding to $6400$ test vectors $\bba^{(k)}$ or $\bbb^{(k)}$, each of size $M=20$. As discussed previously, owing to a {sigmoid} non-linearity on the neural network output, all values are limited to the range~$[0,1]$.

This setup illustrates a transition of the learned variables across a sequence of layers. 
As shown, the values of $\bba^{(k)}$ are mainly concentrated at the lower end of the range in the first and third layers with a major shift in the final layer where it peaks at values close to $1$. 
On the other hand, $\bbb^{(k)}$ values decrease across layers with the peak at the lower end increasing in height as one goes to deeper layers with a majority of $\bbb^{(5)}$ values converging near $0$ on the fifth layer. 
Notice that this validates our result in Theorem~\ref{T:convergence_necessary} and the succeeding discussion. 
It becomes apparent that the proposed architecture leverages the trainable values of $\bba^{(k)}$ and $\bbb^{(k)}$ to expedite convergence in the first few layers and then resembles the WMMSE update rule by learning $\bba^{(k)} = \mathbf{1}$ and $\bbb^{(k)} = \bb0$ in the final layers.  

\subsection{Generalization to variations in network density and size}\label{Ss:spat_den}

\begin{figure*}
	\centering
	\subfigure[]{
			\centering
			\includegraphics[width=0.30\textwidth]{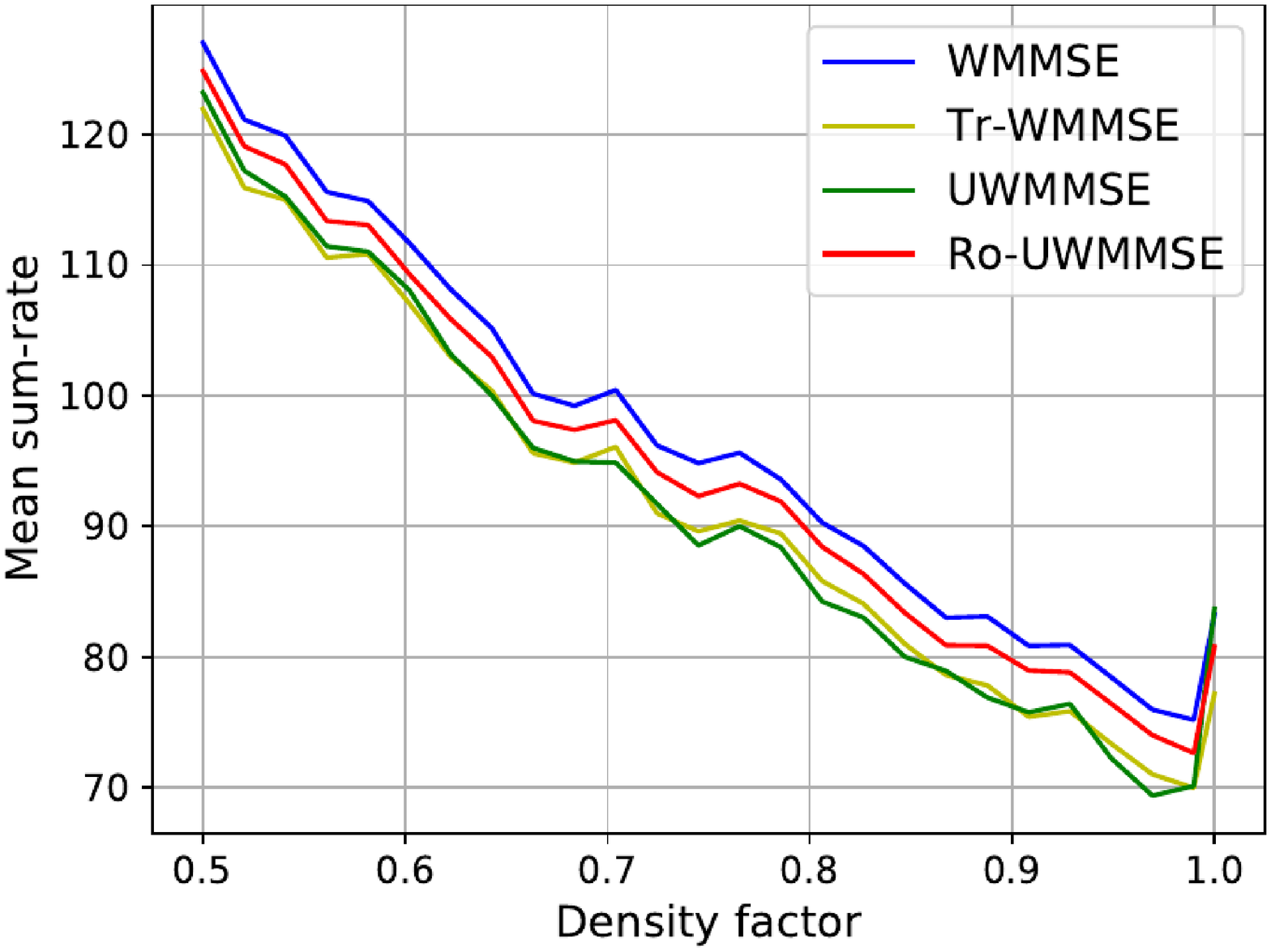}
			\label{Fig:den_left}
		}	
	\subfigure[]{
			\centering
			\includegraphics[width=0.30\textwidth]{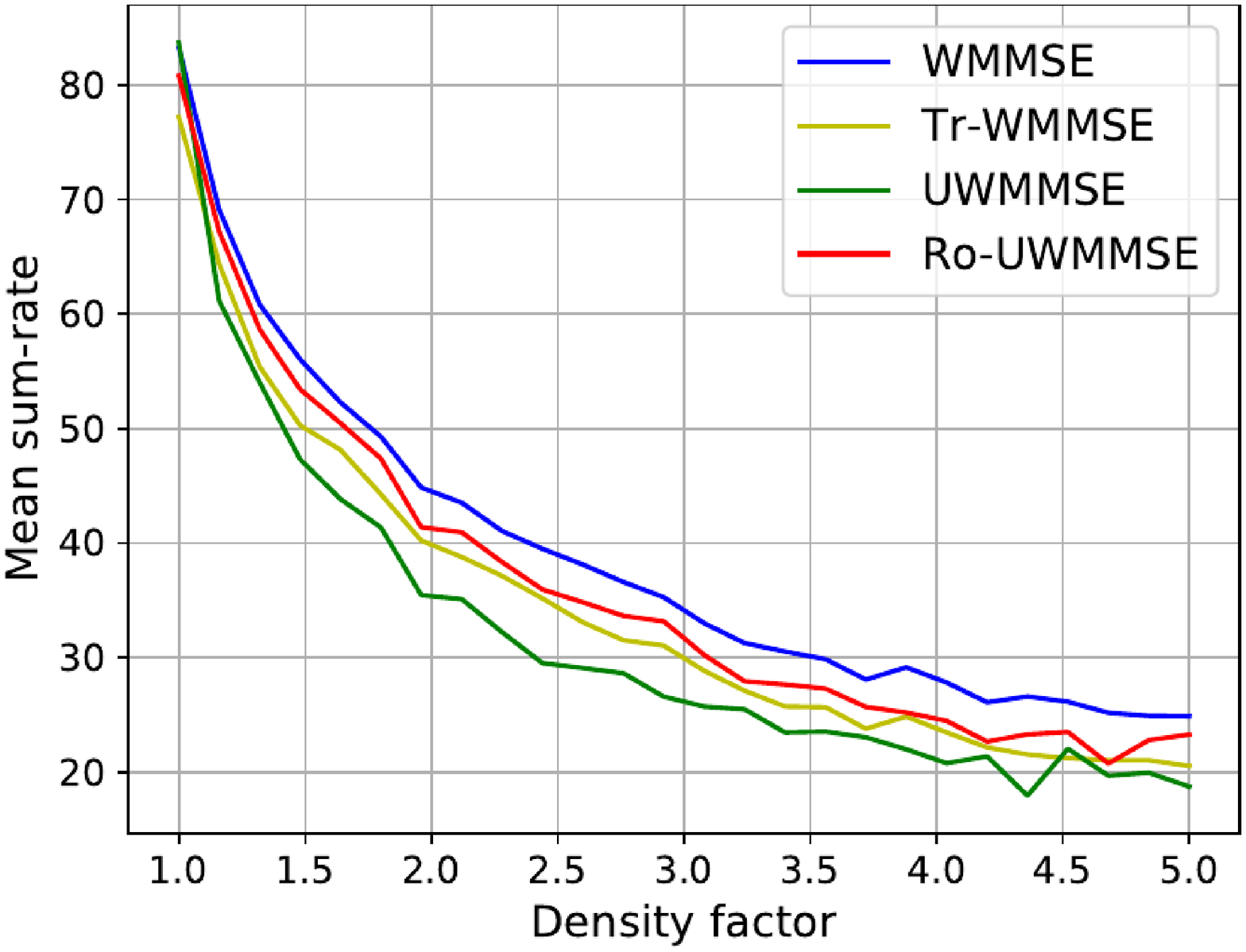}
			\label{Fig:den_right}
		}	
	\subfigure[]{
			\centering
			\includegraphics[width=0.30\textwidth]{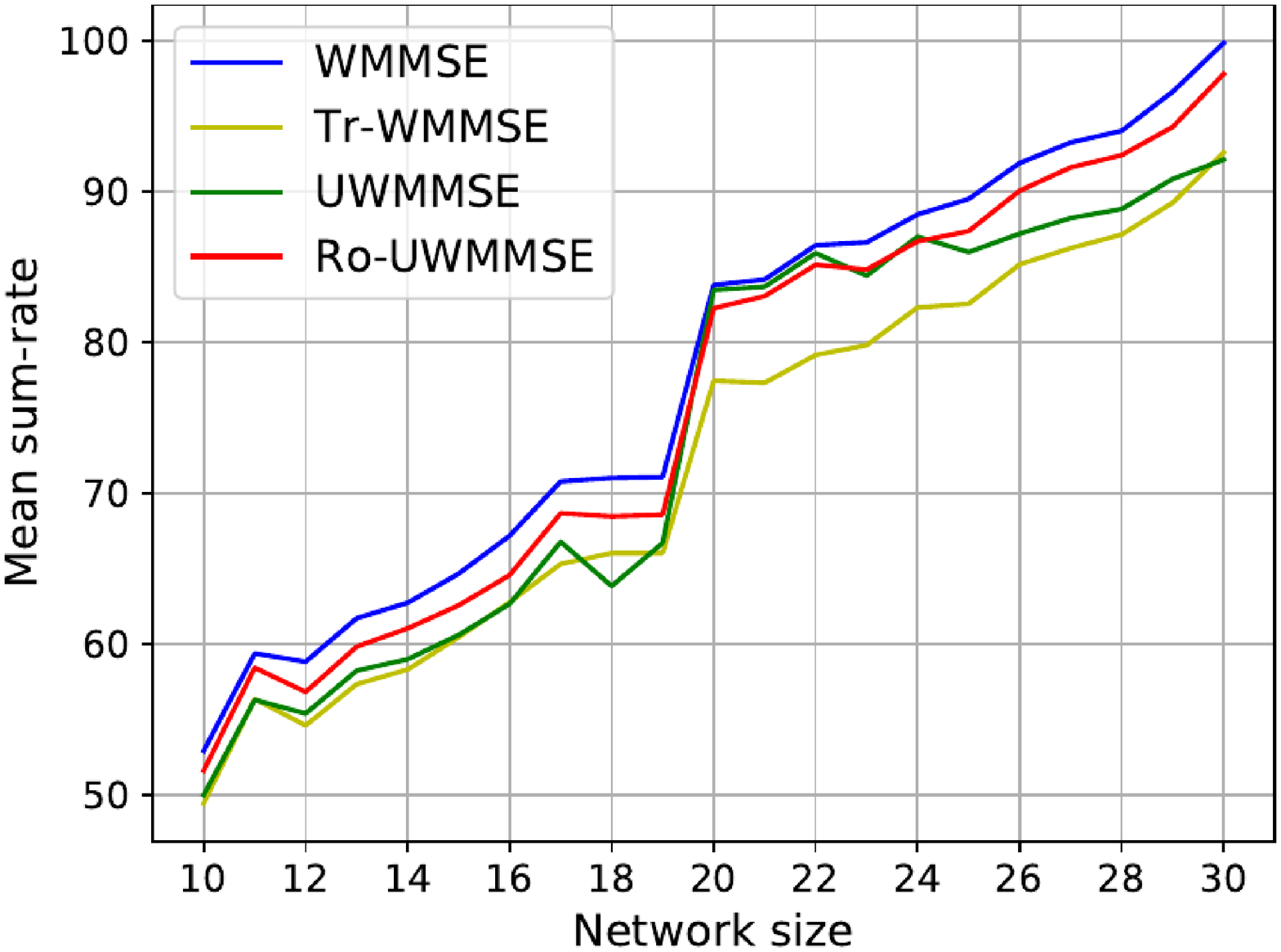}
			\label{Fig:net_size}
		}
		\caption{\small Generalization performance of UWMMSE and a robust version thereof (Ro-UWMMSE) to variations in spatial density and network size compared against WMMSE and Tr-WMMSE baselines. 
		(a)~Mean sum-rate achieved by UWMMSE, Ro-UWMMSE and baselines as density factor $d$ varies in $[0.5,1.0]$, generating sparser networks. 
		(b)~Mean sum-rate achieved by UWMMSE, Ro-UWMMSE and baselines as density factor $d$ varies in $[1.0,5.0]$, generating denser networks. 
		(c)~Mean sum-rate achieved by UWMMSE, Ro-UWMMSE and baselines as network size $M$ varies in $[10,30]$. 
		}
		\label{Fig:variations}
\end{figure*}

So far we have considered a setup in which the underlying topology of a wireless network remains constant across scheduling cycles, while the fading coefficients are sampled randomly at each step. 
However, a more realistic scenario to consider would be one in which the nodes are in motion and thus change their relative positions at each step. 
We simulate this dynamic environment by varying the spatial density of the network. 
We consider two cases. 
One in which nodes move closer to each other resulting in an increase in spatial density and the other in which nodes move further apart resulting in a sparser topology. 
To that end, we define a density factor $d$ that is used to control this setup. 
More precisely, for each value of $d$, a transmitter is dropped at location $\mathbf{t}_i^d = {\bbt_i}/{d}$. 
Its paired receiver $r(i)$, however, is still dropped uniformly at random at location $\mathbf{r}^d_i \in [\bbt_i^d - \frac{M}{4}, \bbt_i^d + \frac{M}{4}]^2$. 
Effectively, the spatial density is varied according to relative positions of the transmitters while maintaining a degree of stochasticity in the receiver positions. 
This is done to create a more realistic and thereby more challenging setting in which the relative distance between a moved transmitter and its receiver is not simply a scaled version of their original distance. 
Further, to simulate both decrease and increase in density, $d$ is varied linearly from $0.5$ to $5.0$. 
We separately plot the variations in the ranges $0.5 \leq d \leq 1.0$ for sparser topologies and $1.0 \leq d \leq 5.0$ for denser topologies in Fig~\ref{Fig:variations}. 

In this experimental setup, we compare UWMMSE trained on a fixed network topology against WMMSE and Tr-WMMSE as baselines. 
Our goal is to analyze the robustness of our proposed approach against variations in the network topology. 
To illustrate this, we introduce a robust version of UWMMSE (Ro-UWMMSE), that is trained on multiple network topologies with varying spatial density. 
At each training step, we generate a batch of independent network topologies as previously described with a randomly sampled $d \in [0.5,5.0]$. 
For each individual topology, we then sample Rayleigh fading coefficients to simulate the channel state information. 
Unlike the baseline UWMMSE that relies on a constant network topology over multiple training cycles, Ro-UWMMSE is regularized in training to allow generalization to multiple network densities. 
In the sparser setting, as evident from Fig.~\ref{Fig:den_left}, there is a constant but moderate gap in performance between WMMSE and UWMMSE, while its robust version follows WMMSE performance closely at almost all points. This is a slightly simpler scenario as the relative distance between transmitters increase, thus, limiting the effect of interference. 
The effectiveness of Ro-UWMMSE is even more pronounced in the contrasting scenario where spatial density increases progressively, bringing the transmitters closer together and, thus, enhancing the effect of interference. 
In this more challenging setting, UWMMSE trained on $d=1.0$ performs poorly on higher $d$ values, lagging even behind the truncated version of WMMSE; see Fig.~\ref{Fig:den_right}. 
However, Ro-UWMMSE is still able to follow the WMMSE performance closely, illustrating the generalization capacity of our unrolling scheme under regularized training. 

In addition to variation in spatial positions, wireless networks also undergo changes in size as nodes are either added or removed from the network. 
To study the performance of UWMMSE in such a setting, we simulate a channel model that undergoes random insertions and deletions of nodes. 
To that end, in each scheduling cycle, a set of transmitters and their corresponding receivers are either removed from the network or a new set of transceivers are added to the network. 
It is important to note that the new transmitters and receivers are still added in the interval $\bbt_i = [-N, N]^2$ for all $i > N$, where $N$ is the original network size. 
In this way, this experiment does not enforce any expansion or contraction of the overall area of the multi-sized topologies. 
Corresponding receivers are dropped $\bbr_i \in [\bbt_i - \frac{N}{4}, \bbt_i + \frac{N}{4}]^2$. 
Rayleigh fading coefficients are sampled independently for each individual topology to construct the CSI matrix.
In this experiment, we train the UWMMSE model on a network of fixed size $N = 20$. 
We then evaluate its performance on networks of size $M \in [10,30]$ by either removing nodes from the original network ($M < N$) or adding new unseen nodes to it ($M>N$). 
Similar to the previous experiment, we compare our method against WMMSE and Tr-WMMSE as baselines. 
Further, to make UWMMSE robust against these variations, we train Ro-UWMMSE on networks of multiple sizes by randomly generating a batch of networks of size $M \in [10,30]$ at each training step. 

As shown in Fig.~\ref{Fig:net_size}, there is a steady drop in UWMMSE performance compared to the WMMSE baseline for both settings $M < N$ and $M > N$, while the best performance is achieved at $M=N$. 
This is not entirely unexpected as in either direction, the underlying topology of the network undergoes significant changes. 
While new nodes that are added to the graph for $M > N$ form hitherto unseen interference patterns, removing a set of transceivers $M < N$ and re-drawing the positions of the receivers for the remaining transmitters also disrupts the existing interference patterns. 
However, Ro-UWMMSE is still able to maintain a performance that is close to WMMSE and, thus, illustrates the generalization capacity of our method to variations in network size under regularized training.

\section{Conclusions}\label{S:Conclusions}

We proposed UWMMSE, a novel neural network approach to solve the problem of power allocation in wireless networks.
The layered architecture of the presented method was derived from the algorithmic unfolding of the classical WMMSE, thus, UWMMSE naturally incorporates domain-specific elements augmented by trainable components.
These components are parameterized by GNNs to account for and leverage the inherent graph representation of communication networks.

Revisiting the motivating question in Section~\ref{S:Modeling}, we have demonstrated that UWMMSE yields solutions that are: 
i)~\emph{Effective}, since the performance attained is comparable with well-established benchmarks;
ii)~\emph{Distributed}, since the layered architecture based on GNNs is amenable to a decentralized implementation; and 
iii)~\emph{Efficient}, since, once trained, a forward pass on the neural architecture requires less time and computation than model-based methods.

A natural extension of our proposed method would be to obtain efficient power allocation in case of complex-valued signals and channels. Additionally, this work opens the door to multiple exciting avenues for future research. Firstly, alternative ways of incorporating learning components to the classical WMMSE iterations can be studied, putting special emphasis in understanding -- theoretically and empirically -- the trade-off between computation and performance for varying number of trainable weights.
Secondly, we can pursue the application of unfolded solutions as the one here developed for other resource allocation problems in wireless communications.
Finally, for cases where security is a concern upon deployment, a relevant direction is the development of architectures that are well-suited for power allocation with missing, noisy, or even adversarial channel information.

\bibliographystyle{IEEEtran}
%
\bibliography{citations}

\end{document}